\documentclass[superscriptaddress,amsmath,amssymb,floatfix,aps,prd,nofootinbib,reprint]{revtex4-1}
\usepackage{natbib}
\usepackage{graphics}
\usepackage{multirow}
\usepackage{amsbsy}
\usepackage{footmisc}
\usepackage{amsmath}
\usepackage{comment}
\usepackage{verbatim}
\usepackage{lineno}

\usepackage{graphicx}
\usepackage{epsf}
\usepackage{subfigure}
\usepackage{color}
\usepackage{threeparttable}
\usepackage{epsfig}
\usepackage{xspace}
\usepackage{enumitem}
\usepackage{hyperref}
\usepackage{ulem}
\usepackage{courier}
\usepackage{appendix}

\usepackage{tabularx}
\newcolumntype{L}[1]{>{\raggedright\arraybackslash}p{#1}}
\newcolumntype{C}[1]{>{\centering\arraybackslash}p{#1}}
\newcolumntype{R}[1]{>{\raggedleft\arraybackslash}p{#1}}

\usepackage[usenames,dvipsnames,svgnames]{xcolor}
\usepackage{latexsym}
\newcommand{\refsec}[1]{Sec.~\ref{sec:#1}}
\newcommand{\reffig}[1]{Fig.~\ref{fig:#1}}
\newcommand{\refeqn}[1]{Eq.~\ref{eqn:#1}}
\newcommand{\reftab}[1]{Table~\ref{tab:#1}}

\DeclareGraphicsExtensions{.jpg,.pdf,.pdf,.eps,.ps}

\hypersetup{
  colorlinks = true,
  urlcolor = black,
  linkcolor=black,
  citecolor=black
}

\usepackage[applemac]{inputenc}

\def\Ad{A_\mathrm{d}}

\def\As{A_\mathrm{sync}}

\def\Bd{\beta_\mathrm{d}}
\def\Bs{\beta_\mathrm{s}}

\def\ad{\alpha_\mathrm{d}}
\def\as{\alpha_\mathrm{s}}

\def\bicep{{\sc Bicep}}

\def\bicepone{{\sc Bicep1}}
\def\biceptwo{{\sc Bicep2}}

\def\keck{{\it Keck}}
\def\keckarray{{\it Keck Array}}
\def\planck{{\it Planck}}

\def\wmap{WMAP}
\def\spt{{\sc SPT}}

\def\sptpol{SPTpol}

\def\bk{\bicep/\keck}

\def\healpix{{\tt Healpix}}
\def\lenspix{{\tt LensPix}}

\def\uksq{$\mu{\mathrm K^2}$}
\def\deg{^\circ}

\def\lcdm{$\Lambda$CDM}

\def\bhat{\ensuremath{\hat{B}^{\mathrm{lens}}}}
\def\beq{\begin{equation}}
\def\eeq{\end{equation}}
\def\beqn{\begin{eqnarray}}
\def\eeqn{\end{eqnarray}}
\def\bl{\pmb{\ell}}

\def\ublu{\bl'}
\def\cii{$C_{L}^{\rm II}$}
\def\ciphi{$C_{L}^{\text{I}\phi'}$}
\def\bk{\bicep/\keck}

\newcommand{\baselineRmarg}{\ensuremath{0.027^{+0.023}_{-0.022}}} 
\newcommand{\baselineRupperlim}{\ensuremath{0.082}}

\newcommand{\baselineoldRmarg}{\ensuremath{0.028^{+0.026}_{-0.025}}}
\newcommand{\baselineoldRupperlim}{\ensuremath{0.090}}

\newcommand{\baselineAdmarg}{\ensuremath{4.2^{+1.1}_{-0.9}}}

\newcommand{\baselineAsupperlim}{\ensuremath{3.7}}

\newcommand{\alensRmarg}{\ensuremath{0.025^{+0.023}_{-0.022}}} 
\newcommand{\alensRupperlim}{\ensuremath{0.081}} 
\newcommand{\alensALmarg}{\ensuremath{1.03 \pm 0.10}} 

\newcommand{\alensoldRmarg}{\ensuremath{0.009^{+0.031}_{-0.009}}} 
\newcommand{\alensoldRupperlim}{\ensuremath{0.079}}  
\newcommand{\alensoldALmarg}{\ensuremath{1.21 \pm 0.17}} 

\newcommand{\simbksr}{\ensuremath{0.024}}
\newcommand{\simbkltsr}{\ensuremath{0.022}} 
\newcommand{\simbkperfectltsr}{\ensuremath{0.018}} 

\newcommand{\oldepsRmarg}{\ensuremath{0.043^{+0.031}_{-0.028}}}
\newcommand{\epsRmarg}{\ensuremath{0.038^{+0.029}_{-0.024}}}

\begin{document}
\title{A Demonstration of Improved Constraints on Primordial Gravitational Waves with Delensing}

\author{The \bicep/\keck\ and SPTpol Collaborations: P.~A.~R.~Ade}
\affiliation{Cardiff University, Cardiff CF10 3XQ, United Kingdom}
\author{Z.~Ahmed}
\affiliation{SLAC National Accelerator Laboratory, 2575 Sand Hill Road, Menlo Park, CA 94025}
\affiliation{Kavli Institute for Particle Astrophysics and Cosmology, Stanford University, 452 Lomita Mall, Stanford, CA 94305}
\author{M.~Amiri}
\affiliation{Department of Physics and Astronomy, University of British Columbia, Vancouver, British Columbia, V6T 1Z1, Canada}
\author{A.~J.~Anderson}
\affiliation{Fermi National Accelerator Laboratory, MS209, P.O. Box 500, Batavia, IL 60510}
\author{J.~E.~Austermann}
\affiliation{NIST Quantum Devices Group, 325 Broadway Mailcode 817.03, Boulder, CO, USA 80305}
\affiliation{Department of Physics, University of Colorado, Boulder, CO, USA 80309}
\author{J.~S.~Avva}
\affiliation{Department of Physics, University of California, Berkeley, CA, USA 94720}
\author{D.~Barkats}
\affiliation{Harvard-Smithsonian Center for Astrophysics, 60 Garden Street, Cambridge, MA, USA 02138}
\author{R.~Basu Thakur}
\affiliation{California Institute of Technology, MS 249-17, 1216 E. California Blvd., Pasadena, CA, USA 91125}
\author{J.~A.~Beall}
\affiliation{NIST Quantum Devices Group, 325 Broadway Mailcode 817.03, Boulder, CO, USA 80305}
\author{A.~N.~Bender}
\affiliation{High Energy Physics Division, Argonne National Laboratory, 9700 S. Cass Avenue, Argonne, IL, USA 60439}
\affiliation{Kavli Institute for Cosmological Physics, University of Chicago, 5640 South Ellis Avenue, Chicago, IL, USA 60637}
% ...

\date{\today}

\begin{abstract}
We present a constraint on the tensor-to-scalar ratio, $r$, derived from measurements of
cosmic microwave background (CMB) polarization $B$-modes with ``delensing,'' 
whereby the uncertainty on  $r$ contributed by the sample variance of the gravitational lensing 
$B$-modes is reduced by cross-correlating against a lensing $B$-mode template.
This template is constructed by combining an estimate of the polarized CMB 
with a tracer of the projected large-scale structure. 
The large-scale-structure tracer used is a map of the cosmic infrared background derived
from \planck\ satellite data, while the polarized CMB map comes from a combination of 
South Pole Telescope, \bk, and \planck\ data.
We expand the \bk\ likelihood analysis framework to accept a lensing template and
apply it to the \bk\ data set collected through 2014 using the
same parametric foreground modelling as in the previous analysis.
From simulations, we find that the uncertainty on $r$ is reduced by $\sim10\%$,
from $\sigma(r)$= \simbksr\ to \simbkltsr, 
which can be compared with a $\sim26\%$ reduction 
obtained when using a perfect lensing template or if there were zero lensing $B$-modes.
Applying the technique to the real data, the  constraint on $r$ is improved from
$r_{0.05} <  \baselineoldRupperlim$ to $r_{0.05} < \baselineRupperlim$ (95\% C.L.).
This is the first demonstration of improvement in an $r$ constraint through delensing.
\end{abstract}
\maketitle

\section{Introduction}
Inflation describes a period of near-exponential expansion during the earliest moments
of the universe.
The inflationary paradigm provides conceptual solutions to problems arising from
the Big Bang description of the early universe including the horizon problem and the flatness problem.
Furthermore, inflationary models make testable predictions about perturbations 
away from perfect homogeneity and isotropy~\cite{kamionkowski2016}.
These predictions have been confirmed in observations of the cosmic microwave
background (CMB) temperature and polarization anisotropies.
They include the Gaussianity, phase-synchronicity, and near-scale-invariance
of the scalar density fluctuations, and super-horizon correlation of the CMB
anisotropies~\cite{planck2018inflation}.
However, one prediction from inflation that has yet to be confirmed 
is the existence of a stochastic primordial gravitational wave (PGW) background.

PGWs are generically predicted in many inflationary models.
Their amplitude is parametrized by $r$, the ratio
of the amplitudes of the tensor and scalar perturbation spectra at a pivot scale 
($k_* = 0.05$~Mpc$^{-1}$ in this work).
If PGWs exist, they would imprint a specific divergence-free ($B$-mode)
signature in the polarization of the CMB~\cite{seljak97, kamion97}.
This makes CMB polarization a promising avenue in the search for PGWs.

However, PGWs are not the only source of $B$-modes.
Thermal dust and synchrotron emission within our galaxy produce
polarized foreground patterns which contain $B$-modes~\cite{planckintXXX,spass18}.
Additionally, there is a source of $B$-modes, called the ``lensing $B$-mode,"
produced by gravitational lensing of the CMB~\cite{lewis2006}. 
If there were no inhomogeneities in the matter between us and the last scattering surface,
then scalar perturbations from inflation would produce a purely curl-free ($E$-mode)
CMB polarization pattern.
However, during their propagation to us, the polarized CMB photons undergo 
small gravitational deflections by the forming large-scale structure 
along the line of sight.
This produces a $B$-mode component which is small compared to the source $E$-modes,
and which has already been detected
by a number of experiments~\cite{Hanson:2013hsb, polarbearbb14, keisler15, louis16,bk6,polarbearbb17,bk10}.

The \bk\ experiments have deployed CMB polarization telescopes optimized for measurements at the
``recombination bump" in the predicted PGW-generated B-mode spectrum
(harmonic multipoles $\ell \sim 80$, or angular scales of $\sim 2$~degrees). 
To separate out the Galactic dust and synchrotron components, 
which have different frequency spectral shapes than the black-body emission of the CMB,
\bk\ observes in several frequency bands, and the $r$ analyses also
incorporate maps at additional frequencies from the \wmap\ and \planck\ satellites.
The existing analysis pipeline takes all possible auto- and cross-spectra of the maps
at different frequencies and compares these against a parametric model
of CMB and foregrounds~\cite{bk6,bk10} to set constraints on $r$ which are
close to optimal given the available data.
Alternative approaches involving ``cleaning coefficient'' subtraction of
a dust template map (as measured at a higher frequency)
would in general be less powerful~\cite[e.g.][]{bkp}.

In contrast to the foregrounds, the lensing component has the same frequency spectral shape
as the PGW component, and thus cannot be constrained using multi-frequency observations.
Given an estimate of the projected gravitational potential responsible for CMB lensing 
and the observed CMB $E$-mode pattern, 
one can estimate the $B$-modes which have been produced by the lensing effect.
Subtracting these from the observed $B$-modes
has been demonstrated to reduce
$B$-mode power in several recent works~\cite{manzotti17, carron17, plancklens18, pbdelens19, actdelens20}. 
However, none of these works have demonstrated
a reduction in the $B$-mode measurement uncertainties at large angular scales---a 
necessary step to achieve improved constraints on PGWs.
This subtraction process is usually referred to as ``delensing."
But in this work, we take a different approach and therefore broaden the meaning of delensing
to include any process which reduces the effective lensing sample variance
in the $B$-mode measurements. 
Specifically, we extend the \bk\ analysis pipeline to accept an
estimate of the lensing $B$-modes as a ``lensing template''---an
additional pseudo-frequency band against which cross-spectra are taken.
This (optimally) reduces the effective sample variance of the lensing $B$-mode
component, and hence reduces the uncertainty of the PGW contribution.

The lensing potential $\phi$ can be computed using higher-order statistics of the CMB pattern
itself~\cite{huokamoto02}.
However, since the lensing potential is a weighted integral of the mass distribution along
the line of sight between us and the last scattering surface, we may
also approximate it by other tracers of this mass distribution.
At the noise levels of current CMB observations, it turns out
to be better to use a cosmic infrared background (CIB)~\cite{simard15,sherwin15}
map rather than one of the available CMB lensing reconstructions~\cite{plancklens18,wu2019}
directly.
To use an alternate tracer of $\phi$, we need to know the
degree of correlation between it and the true CMB lensing potential---if
this were misestimated it could potentially lead to a false detection of
PGW.
This correlation may be found empirically from the cross-correlation
of the tracer with a reconstruction of the CMB lensing potential.
In this paper, we use a CIB map from \planck\ generated using the
Generalized Needlet Internal Linear Combination
(GNILC) component separation algorithm~\cite{planck2015cibdust}
as the $\phi$ tracer and estimate its correlation with the lensing potential using
a \planck\ minimum-variance lensing map~\cite{plancklens15}.

To estimate the lensing template, in addition to the tracer of the lensing potential,
one also needs the best available estimate of the observed CMB
polarization pattern.
Since the lensing operation mixes modes over a wide range of angular scales,
the inclusion of small-scale $E$-modes is important for precise estimation
of the lensing $B$-modes at the angular scales of interest ($\ell \sim 80$).
Therefore, we use arcminute-resolution maps from the South Pole Telescope (SPT)
second-generation camera \sptpol, augmenting these with polarization
measurements from \bk\ and \planck.

In this paper, we add the CIB-derived lensing template to the
previous ``BK14'' analysis~\citep{bk6}
which utilizes data from \bicep/\keck\ through the 2014 observation season.
With the addition of the lensing template, we demonstrate a $\sim10\%$ reduction in the 
uncertainty on $r$ for the BK14 data set, to be compared with a
$\sim26\%$ reduction in the uncertainty on $r$ when using a perfect lensing template
or if there were zero lensing $B$-modes.
This shows that the lensing sample variance is a subdominant fraction of the uncertainty on $r$
for BK14.
However, it will be an increasingly limiting factor going forward.
Therefore, this analysis serves as a proof of principle, and a first step towards
future analyses where delensing will more significantly improve $\sigma(r)$.

This paper is organized as follows:
In \refsec{method}, we describe the construction of the lensing template
and the extension to the \bk\ pipeline to include the lensing template.
In \refsec{datasims},
we describe the data and simulation sets of the CMB maps,
how we combine the $Q/U$ maps from \sptpol, \bk, and \planck, 
and the data and simulations of the $\phi$ tracer.
We validate our simulations and pipeline in \refsec{simval} and test for
systematics in  \refsec{syscheck}. 
We present our results in \refsec{results} and conclude in \refsec{conclusion}.

\section{Method}\label{sec:method}
\begin{figure*}
\begin{center}
\includegraphics[width=0.98\textwidth]{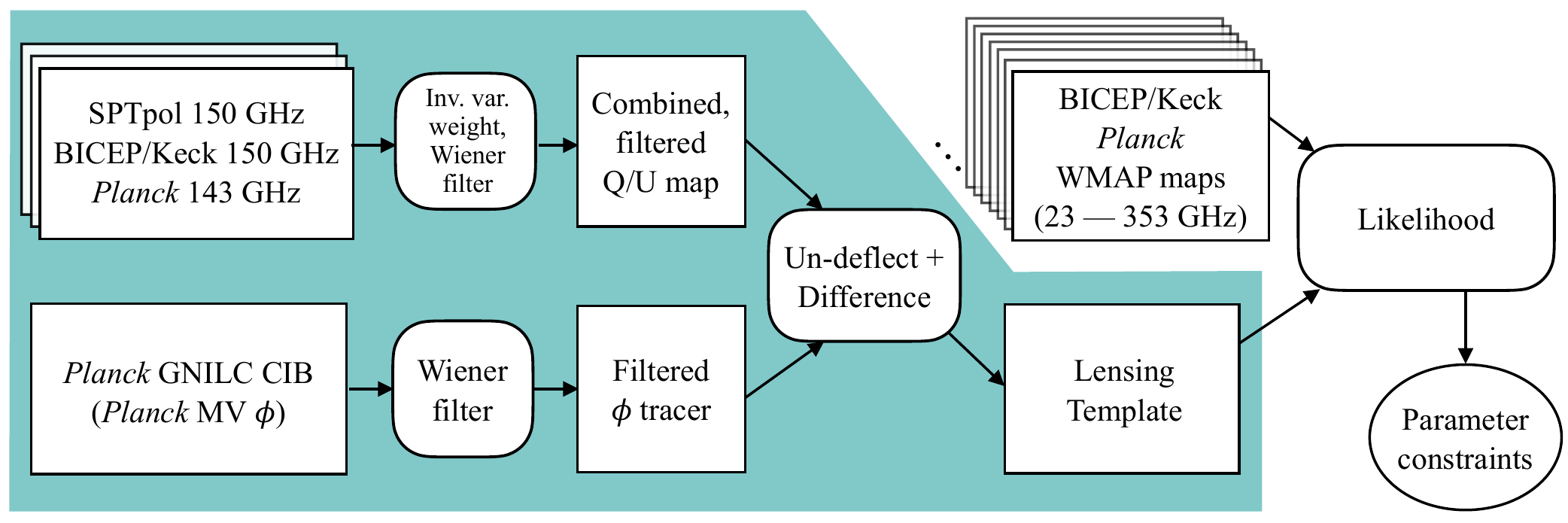}
\end{center}
\caption{Schematic of the analysis flow in this work. The rectangular blocks denote input maps;
the blocks with rounded corners denote operations on maps.
The teal-colored region highlights the inputs to, and processes involved in generating, the lensing template.
The input maps include the \sptpol, \bk, and \planck\ 
$Q/U$ maps, the \planck\ GNILC CIB map, and the \planck\ minimum-variance (MV) reconstruction of $\phi$. 
The \planck\ MV $\phi$ is in parentheses because, instead of using it as a $\phi$ tracer, we use it to filter and
normalize the CIB map and for generating simulations.
The unshaded region denotes the standard \bk\ $r$ analysis, where auto- and cross-spectra of
multi-frequency maps from \bk, \planck, and \wmap\ form the input data for computing
likelihoods to extract parameter constraints.
The lensing template is injected into the standard analysis as an additional pseudo-frequency band.
}
\label{fig:analysis_flowdown}
\end{figure*}

In this section, we describe new elements added to the \bicep/\keck\ analysis framework
to incorporate information on the lensing $B$-modes in the \bicep/\keck\ patch, 
with the aim of reducing the effective uncertainty of the observed $B$-modes,
and thereby reducing the uncertainty on $r$.
We illustrate the incorporation of the lensing template into the \bk\ likelihood
analysis framework schematically in~\reffig{analysis_flowdown}.
There are two main areas of new development:
(1) constructing a lensing template, and
(2) extending the \bicep/\keck\ pipeline to include the lensing template. 
We will describe each aspect in the following subsections.

\subsection{Constructing the lensing template}\label{sec:construct_lt}
The key element to constraining the lensing $B$-modes in the \bicep/\keck\ patch is making an
estimate of these modes. 
To do this, we use two inputs:  
(1) a tracer of the CMB lensing potential $\phi$ from large-scale structure observations
and (2) observed $Q/U$ polarization maps.
We construct the lensing template using an ``undeflect-and-difference" method in which
we undeflect the observed $Q/U$ maps using the $\phi$ tracer and subtract the
undeflected maps from the input.

Formally, we take the lensed, polarized CMB fields $\tilde{X}$, which are related to the unlensed
CMB fields $X$ by
\begin{equation} \label{eqn:lensing}
\tilde{X}_{\pm}(\hat{n}) = X_{\pm}(\hat{n} + \nabla \phi(\hat{n}))
\end{equation}
where $X_{\pm} = (Q \pm i  U)$ and $\nabla\phi$ denotes the deflection field~\cite{hu2000}. 
We undo the deflection by remapping the $\tilde{Q}$ and $\tilde{U}$ polarization fields
by $- \nabla \phi$, evaluated at the lensed positions $\hat{n}'$ (not delensed positions $\hat{n}$),
\begin{equation}
X^d_\pm (\hat{n}) = \tilde{X}_\pm(\hat{n}'),
\end{equation}  
where $\hat{n}'=\hat{n}-\nabla\phi(\hat{n}')$ and $X^d$ denotes the undeflected field. Therefore,
\begin{equation}
X_\pm(\hat{n})=X^{d}_{\pm}(\hat{n}) = \tilde{X}_{\pm}(\hat{n} - \nabla \phi ({\hat{n}}-\nabla \phi ({\hat{n} - ...}))) 
\end{equation}
where the last expression is used for the practical implementation, 
and $...$ denotes the recursion that locates the position to which 
the value at $\hat{n}$ was deflected from the unlensed plane.

Specifically, the undeflection is implemented by first computing the amount of
deflection at the lensed positions, denoted by ($dx,dy$), on the delensed map
pixel grid $(x,y)$.
To do that we first evaluate $\nabla \phi$ at $(x,y)$ to get ($dx',dy'$), 
and then we evaluate $\nabla \phi$ at $(x-dx',y-dy')$, and so on.
We find that the solution converges after 1 recursion, which means that
with the notation given, ($dx,dy$) is $\nabla \phi$ at $(x-dx',y-dy')$.
The evaluation of $\nabla \phi$ at any grid point is done by interpolating 
$\nabla \phi$ values in {\sc Healpix} format using first-order Taylor expansion.
We then remap the $\tilde{Q}/\tilde{U}$ map pixels at ($x-dx, y-dy$) to ($x,y$)
via cubic interpolation.
We note that by evaluating the deflection field at the lensed positions, 
we do not incur the small $\mathcal{O}(\nabla \phi \cdot \nabla)\nabla \phi$ error
found in similar algorithms that evaluate $\nabla\phi$ at the delensed positions~\cite{anderes15, green16, carron17}.

The lensing templates $Q^t/U^t$ are then derived by subtracting the obtained
undeflected map from the observed (lensed) one:
\begin{align}
Q^t (\hat{n}) &= \tilde{Q} (\hat{n}) - Q^d (\hat{n})\\
U^t (\hat{n}) &= \tilde{U} (\hat{n}) - U^d (\hat{n}).
\end{align}
We test the algorithm on noiseless lensed simulations
using the $\phi$ maps which were used to lens them.
The correlation of the resulting lensing $B$-mode
template with the difference of the lensed and unlensed input skies is 
$\gtrsim95\%$ for the angular scales used in this analysis. 
This is sufficiently accurate at the current noise levels.
In other work, lensing templates have also been constructed after transforming
to harmonic space, converting to $E/B$, and
lensing by a $\phi$ tracer using expressions derived from the first-order
Taylor expansion of \refeqn{lensing}~\cite{plancklens18, manzotti17, pbdelens19}. 
At the noise levels of the current analysis, the two approaches perform similarly in constraining 
the lensing $B$-mode contribution to the observed $B$-modes.
We discuss in more detail the differences of the two approaches in Appendix~\ref{app:templates}.

Since the undeflect-and-difference operation corresponds to an all-with-all mixing in
Fourier space, to obtain the lowest possible lensing template noise
in the $\ell$ range of interest we first Wiener filter~\citep[e.g.][]{green16}
the $Q/U$ and $\phi$-tracer maps.
We filter the $Q/U$ maps by a 2D Wiener filter in Fourier space
\begin{align}
\label{eqn:qu_wiener}
\tilde{Q}(\bl) &\rightarrow  \frac{C_{\bl}^{EE}}{C_{\bl}^{EE} + N_{\bl}^{EE}} \, \tilde{Q} (\bl), \\
\tilde{U}(\bl) &\rightarrow  \frac{C_{\bl}^{EE}}{C_{\bl}^{EE} + N_{\bl}^{EE}} \, \tilde{U}(\bl), 
\end{align}
to account for anisotropic noise and mode-loss due to filtering.
$C_{\bl}^{EE}$ and $N_{\bl}^{EE}$ are 2D power spectra of the $E$-mode
signal and noise components, 
constructed from 
a weighted combination of $Q/U$ maps from the
three experiments \sptpol, \bk, and \planck. 
We describe the procedure to combine the $Q/U$ maps 
and the details of the Wiener filter in~\refsec{combine_qu}.

We filter and normalize the $\phi$ tracer in spherical harmonic space according to
\beq\label{eqn:tracer_wiener}
\phi^{\rm \cal{T}}_{LM}  =
\left( \frac{C_{L}^{\rm \cal{T}\phi'}}{C_{L}^{\rm \cal{T}\cal{T}}} \right) {\cal T}_{LM},
\eeq
where $\cal{T}$ denotes the tracer
and $\phi'$ is an unbiased, but noisy map of the true CMB lensing potential~\citep{simard15,sherwin15}.
To see that this weighting is a joint normalization and Wiener filter, we write
 ${\rm \cal{T}}_{LM} = g_L \phi_{LM} + n_{LM}$, where $g_L$ is the relative normalization factor
(and unit conversion), $\phi$ is the true (noiseless) lensing potential, and $n$
is the effective noise in the tracer pattern
(the part which does not correlate with $\phi$) with power
spectrum $N_L^{\cal{TT}}$.
Expanding and taking the expectation value, we get
\beq\label{eqn:tracer_wiener_norm}
\frac{C_{L}^{\rm \cal{T}\phi'}}{C_{L}^{\rm \cal{T}\cal{T}}} =
\frac{g_LC_L^{\phi\phi}}{g_L^2C_L^{\phi\phi} + N_L^{\rm \cal{TT}}} =
g_L^{-1}\frac{C_L^{\phi\phi}}{C_L^{\phi\phi} + N_L^{\rm \cal{TT}}/g_L^2},
\eeq
fulfilling its role of normalization and filtering. 
In this paper the tracer $\cal{T}$ is a CIB
map from \planck\ and $\phi'$ is a lensing reconstruction map also from
\planck---see~\refsec{cib_maps} below for further details.

With the $Q/U$ lensing templates constructed, we then take them as
an additional pseudo-frequency band for input into the existing \bk\ analysis.

\subsection{Adding the lensing template to the existing analysis framework}\label{sec:add_lt}

The development of the existing \bk\ $r$ analysis framework has been described in a series
of papers~\cite{bk1,bkp,bk6,bk10}.
Briefly, we take all possible auto- and cross-power spectra
between the available frequency bands, and then compare the resulting
set of bandpowers to their expectation values under a parametric model
of lensed-\lcdm+dust+synchrotron+$r$ using an expansion of the Hamimeche-Lewis
likelihood approximation~\cite{hl2008}.
It is a straightforward extension to this framework to include the lensing
template as an additional pseudo-frequency band.
To do this we require reliable simulations of the signal
and noise content of the lensing template so that we can (1) debias
its auto-spectrum, (2) determine the expectation values
of the auto- and cross-spectra involving the lensing template, and
(3) determine the variance of these bandpowers, and their covariance
with other bandpowers.
These simulations are described in \refsec{datasims} below.
Here we describe a few complications with respect to the
normal procedure which arise in the steps above.

The lensing template is formed from two kinds of input
maps (the $Q/U$ maps and the $\phi$ tracer) which both contain relevant amounts of noise.
The \planck\ CIB map has very high signal-to-instrumental-noise.
However, the integrated dust emission from star-forming galaxies back
to the last scattering surface weights differently over redshift
than the deflection of CMB photons, and these galaxies do not
perfectly trace the underlying mass density field.
This means that the CIB only partially correlates with the true lensing potential.
For the purposes of this paper, the $\phi$ tracer signal is the
portion of the CIB that is correlated with the true lensing potential $\phi$; 
the $\phi$ tracer noise corresponds to the uncorrelated portion. 
We detail our $\phi$ tracer simulations in~\refsec{cib_sims}.

We remove the noise bias of the lensing template auto-spectrum by subtracting the noise auto-spectrum
estimated from simulations.
Schematically, the lensing template $B$-mode auto-spectrum is
\beq
\label{eqn:temp_spec}
\begin{split}
\left< L_B^2 \right> = & \left< ((s_{QU} + n_{QU}) \ast ( s_\phi + n_\phi ))^2 \right> \\
		      = & \left< (s_{QU} \ast s_\phi)^2 \right> + \left< (s_{QU} \ast n_\phi)^2 \right> + \\
                       & \left< (n_{QU} \ast s_\phi)^2 \right> + \left< (n_{QU} \ast n_\phi)^2 \right>,
\end{split}
\eeq
where $s_X$ and $n_X$ denote the signal and noise components of field $X \in [QU, \phi]$ 
and \(\ast\) denotes the following steps: undeflect-and-difference,
Fourier transform, and convert from $Q/U$ to $B$-modes.
In writing the second line, we have assumed
all the cross terms have zero expectation value.
We estimate the noise auto-spectrum from simulations as
\begin{equation}
\label{eqn:lt_noise_spec}
\left< (n_{QU} \ast (s_\phi + n_\phi))^2 \right> + \left< (s_{QU} \ast n_\phi)^2 \right>
\end{equation}
averaged over all simulation realizations and subtract it from~\refeqn{temp_spec}.
The $Q/U$ and $\phi$ input signal and noise maps are Wiener filtered
in the same way as the data maps (and the simulation signal+noise maps).
Empirically, when adding this inferred noise bias to the mean of the
signal-only simulation spectra
($Q/U$ signal undeflected with $\phi$ tracer signal) 
one obtains the mean of the signal+noise simulation spectra
($Q/U$ signal+noise undeflected with $\phi$ tracer signal+noise)
to high fractional precision.
 
In the \bk\ standard procedure, the filter/beam suppression of the bandpower values
is computed using 
sets of maps which each contains power at only a single multipole $\ell$
passed through the ``observing matrix''
as described in Sec.~VI.C of~\cite{bk1}.
However, since the lensing template is derived in a very different manner
to the standard \bk\ maps, the usual observing matrix is not applicable,
and we fall back to a simulation-based approach.
We rescale both the data and simulation lensing template
auto- and cross-spectra by the ratio of the input lensing spectrum $C_{\ell}^{BB}$
to the average of the signal-only simulation bandpowers.
This step overrides the normalization part of~\refeqn{tracer_wiener_norm} applied to the $\phi$ tracer.
However, accurate knowledge of the degree of correlation between the lensing
tracer and the true lensing potential is still required to avoid
bias on $r$ (see~\refsec{simval}).

In the standard \bk\ procedure, the bandpower covariance matrix is constructed by
taking the auto- and cross-spectra of the signal and noise components
of the simulations as described in Appendix H of~\cite{bk10}.
Since the lensing template is formed from two maps which both have signal
and noise components we expand the usual procedure to form additional
cross spectra and combine the results appropriately.

With this extended analysis framework, we can now incorporate lensing templates
constructed using simulations and data to the \bk\ likelihood and constrain the model parameters. 

\section{Data and Simulations} \label{sec:datasims}
The \bk\ analysis pipeline relies on signal-only, noise-only, and signal+noise
simulations, the construction of which is described in Sec.~V of~\cite{bk1}.
We reuse the data maps and simulations including Gaussian realizations of Galactic dust
from the BK14 analysis unchanged.
The data maps include the \wmap\ and \planck\ bands with \bk\ filtering applied
(as described in Sec. II.A of~\cite{bkp}).
To add the lensing template as an additional pseudo-frequency band, we need
data maps and corresponding simulations of it.
Since the lensing template is constructed from $Q/U$ CMB maps and 
a CIB $\phi$ tracer, we in turn need data maps and simulations of both of these.
As a pre-step, we combine the \sptpol, \bk, and \planck\ $Q/U$ maps to
generate a synthetic map which has the best possible signal-to-noise
at all points in the 2D Fourier plane.
\reffig{analysis_flowdown} gives a schematic view of the steps involved in 
generating the various maps.

\subsection{$Q/U$ CMB maps}
Below we describe the data processing of the \sptpol,  \bk, and \planck\ $Q/U$ maps
that are relevant in the construction of the combined $Q/U$ maps and their Wiener filter.
The combined, Wiener filtered $Q/U$ maps are the inputs to the undeflect-and-difference
step which is used to construct the lensing template.

\subsubsection{Data CMB maps}
\textit{\sptpol\  maps: }
We use \sptpol\ maps made specifically for this analysis using 150~GHz observations
taken between 2013 and 2015 by the \sptpol\ camera~\cite{austermann12} on the
South Pole Telescope~\cite{carlstrom11}.
The \sptpol\ 500 deg$^2$ survey field is centered at RA 0h and Dec. $-57.5\deg$,
matching the \bicep/\keck\ field. 
The polarization map depth is $\sim10\,\mu$K$\,$arcmin in the multipole range of $300 \lesssim \ell \lesssim 2000$.
The time stream processing is identical to that in~\cite{henning18}, 
except for the polynomial-filter order and the low-pass filter. 
We fit and subtract a third-order/sixth-order polynomial 
from the time stream of each detector over the RA extent of the lead-trail/full-field observations.
We choose the low-pass filter based on the pixel size. 
This set of \sptpol\ maps is binned into 5~arcminute-sized pixels, which
are a $\times 3$~resolution superset of the \bk\ map pixels.
To reduce aliasing given the pixel size, we apply a low-pass filter to the time stream
that corresponds to $\ell \sim 1900$. 
The polarization maps, in addition to the calibration factors included through calibrating 
the temperature map against \planck, have an extra polarization calibration factor ($P_{cal}$) applied.
The polarization calibration factor is taken from~\cite{henning18} and is obtained by forming a cross-spectrum between the 
\sptpol\ $E$-mode map and an $E$-mode map from \planck.\footnote{\href{http://pla.esac.esa.int/pla/aio/product-action?MAP.MAP_ID=COM_CMB_IQU-commander_1024_R2.02_full.fits}{\planck\ COMMANDER maps: \\ \texttt{COM\_CMB\_IQU-commander\_1024\_R2.02\_full}}} 
We discuss impacts on $r$ from biases in $P_{cal}$ in~\refsec{sys}.

\textit{\bicep/\keck\ maps:} 
We use the \biceptwo/\keck\ 150~GHz band $Q/U$ maps from BK14.
These have noise of $\sim~3\,\mu$K$\,$arcmin over an effective area of
395 deg$^2$ centered at RA 0h, Dec. $-57.5\deg$.
The \biceptwo\ and \keckarray\ telescopes have $\sim$30~arcminute
resolution at 150~GHz. This limits the highest angular multipole 
to which they are sensitive to $\ell$ of hundreds.
As described in Sec. III \& IV of~\cite{bk1} the construction of the maps
involves time-stream filtering. 
Specifically, a third-order polynomial was subtracted from the time streams of each detector over each scan.
Across the $\sim\,$30$\deg$ scan throw on the sky, this approximately corresponds to removing $\ell_x < 20$ modes.
These maps are binned in 0.25$\deg$ rectangular pixels in RA and Dec, and calibrated by
forming cross-spectra with the \planck\ temperature map.

\textit{ \planck\ maps:}
We use the 143 GHz $Q/U$ ``full mission" maps from \planck\ public release 2 as the
input to the combined three-experiment $Q/U$ maps.
We convert the map to $a_{\ell m}$,
apply an anti-aliasing filter by low-pass filtering at $\ell=2100$,
render a $N_{\rm side}=2048$ \healpix\ map, and 
interpolate to the same 5-arcminute pixel grid as used  for
the \sptpol\ maps.

\subsubsection{Simulated CMB maps}
\label{sec:cmbsims}
We reuse the BK14 simulated maps unchanged.
We thus need to make corresponding simulations for the \sptpol\
and \planck\ maps.
The \bk\ CMB sky realizations have remained the same since originally
described in Sec. V of~\cite{bk1}.
These are the unlensed $a_{\ell m}$, Gaussian realizations of $\phi$ given the input cosmology, 
and lensed $a_{\ell m}$ generated using \lenspix.\footnote{\url{https://cosmologist.info/lenspix/}}
The \bk\ simulations were originally generated with a maximum
$\ell$ of 1536 which is adequate given the beam sizes of the telescopes.
To match more closely the pixel scale of the \sptpol\ and \planck\ maps in this analysis,
we generate additional higher-$\ell$ $a_{\ell m}$,
graft these onto the existing unlensed values, pass through \lenspix, and graft the
output onto the existing lensed values.
Since lensing to some degree mixes angular scales this is clearly only
approximately correct, but we note that the
amount of lensing $B$-modes below $\ell$ of 350 sourced from $E$-modes
between $\ell$ of 1536 and 2100 (the pixel-scale) is negligible~\citep[see e.g.\ Fig.~2 of][]{simard15}.
We refer to this set of input lensed and unlensed $a_{\ell m}$ as the extended set
and the original set as the standard set.

We generate \sptpol\ simulations for this analysis using the extended set of $a_{\ell m}$.
In a procedure similar to that used to generate the existing \bk\ simulations,
we multiply the input $a_{\ell m}$ by the instrument beam,  
``mock-observe'' the skies by creating time-stream samples given the pointing information of each detector,
apply the same time-stream level filters as applied to data,
and bin to maps in the pixelization used for the real data.
Corresponding noise realizations are generated by the standard method
used in both \bk\ and \spt\ analyses---differencing combinations of halves of data maps,
where the halves are defined so that the weights of each half are close to equal. 

We generate simulated maps for \planck\ 143~GHz by first taking the 
$a_{\ell m}$ from the extended set and multiplying them by the \planck\ 143~GHz beam. 
We then low-pass filter and process as for the real \planck\ map.
Corresponding noise realizations are taken from 
the \planck\ FFP8 simulations and processed identically to the real \planck\ map. 
We generate 499 realizations of signal and noise skies for each experiment as is the \bk\ standard.

\subsubsection{Combining and Filtering the $Q/U$ maps from \sptpol, \bk, and \planck}
\label{sec:combine_qu}

A factor that impacts the delensing efficiency 
(the recovery of the lensing $B$-modes) is the per-mode noise of the
input $Q/U$ maps. 
The lower the noise per mode, the better the lensing templates
trace the true lensing $B$-modes.
The lensing $B$-modes at multipole $\ell$ are mostly sourced by $E$-modes 
from a range of multipoles slightly higher in $\ell$ (smaller in angular scale)~\cite[see e.g. Fig.~2 of][]{simard15}. 
\bk\ does not image these smaller-scale $E$-modes very well because of its large beam size.
Therefore, it is advantageous to combine with polarization measurements from 
other, higher-resolution experiments such as \sptpol\ and \planck\
 to increase the signal-to-noise ratio of the input $Q/U$ maps, and thus the $E$-modes.

We combine the three maps in Fourier space.
We divide from the 2D mode sets of the three experiments their respective
2D transfer functions, taken as the square-root of the mean of the 2D
$E$-mode power spectra of the signal-only simulations divided by the mean of the
corresponding spectra of the (unfiltered) input maps.
We also divide the 2D noise power spectra by the
same ratio (without the square root).
We then combine the \sptpol, \bk, and \planck\ $Q/U$ modes
using an inverse-variance weighting taken from the mean of the 2D noise power spectra.
Specifically, the combined $Q/U$ mode sets are
\begin{equation}
{\cal X}(\bl) = \sum_i w_i(\bl) {\cal X}_i(\bl), 
\end{equation}
where ${\cal X} \in [Q, U]$, $i \in $ [\,\sptpol, \bk, \planck \,],  and $w_i$ denotes the weight
\begin{equation}
\label{eqn:noise_weight}
w_i (\bl) = \frac{N^{-1}_i(\bl)}{\sum_i N^{-1}_i(\bl)}.
\end{equation}
Here, $N_i(\bl)$ denotes the mean of the transfer-function-divided
2D angular power spectra of the $E$-mode noise realizations from experiment $i$. 
We additionally impose $\ell_x$ cuts by artificially increasing the noise below 
some $\ell_x$ to remove modes that are empirically found to be unrecoverable
due to the scan-wise timestream filtering. 
We set $\ell_x$ to 25 for \bk\ and $\ell_x$ to 50 for \sptpol.

Before passing the combined $Q/U$ map to the lensing template construction step, 
we apply a Wiener filter as described in \refsec{construct_lt} above.
The $C_{\bl}^{EE}$ in \refeqn{qu_wiener} is the 2D input
$E$-mode power spectrum and the $N_{\bl}^{EE}$ of the combined modes is
\begin{equation}
N^{EE}_{\bl} = \sum_i w^2_i(\bl) N_i(\bl),
\end{equation}
with $w_i(\bl)$ given by~\refeqn{noise_weight}. 

In the above, we transform to Fourier space and back again, 
and hence need to choose an apodization mask.
Due to the small instantaneous field of view of the \sptpol\
camera as compared to the size of the observation region, the
integration time map (inverse noise variance map) is a near
uniform rectangular box tapering to zero over a few degrees at the edges.
In contrast, the \bk\ integration time map has no uniform central region
and tapers smoothly and continuously from a peak in the middle
(see for instance Fig.~1 of~\cite{bkp}), with non-zero coverage
extending well outside the \sptpol\ region.
(\planck\ observes the full sky and has close to uniform coverage
across the sky region in question.)
To perform the map combination we need to pick a single apodization
function for all three input maps.
We choose to use the one built from the \sptpol\ integration time map, 
with a cosine taper with radius of 1 degree. This is because \sptpol\ 
is the experiment
with the most restrictive sky coverage, but the best mode coverage.
This means that the resulting lensing template does not cover
the full \bk\ sky region. 
In addition, because of the chosen spatial weighting of pixels, 
we introduce sub-optimality in the combination.

\begin{figure*}
\begin{center}
\includegraphics[width=\textwidth]{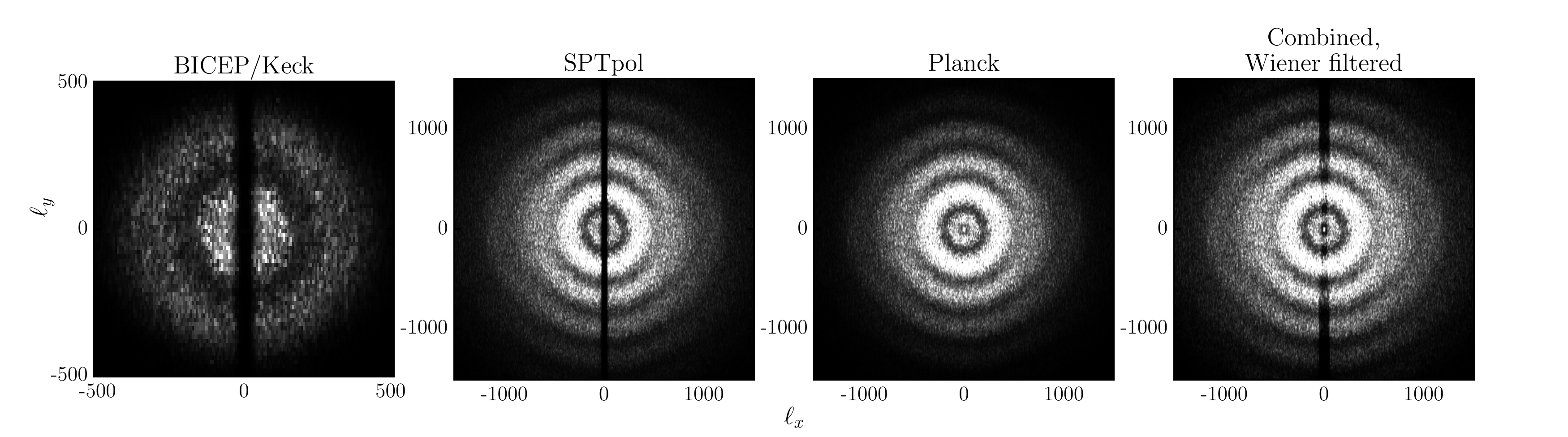}
\end{center}
\caption{Simulated 2D $E$-mode signal power spectra of \bk, \sptpol, and \planck.
The axis scales for the \bk\ Fourier plane are zoomed in compared with the rest 
of the panels to focus on the modes accessible by \bk's small apertures.
The color stretch in all four panels is identical. 
For \bk\ and \sptpol\, because the observations are being made at the
South Pole with scans along the azimuth direction, scan-wise filtering 
leads to modes along the $\ell_y$ axis being suppressed.
These filtered modes along the
$\ell_y$ axis can be partially filled in using measurements from \planck.
To generate the combined, Wiener filtered 2D $E$-mode signal power spectra 
on the rightmost panel, 
the three sets of modes to the left are corrected for beam and filtering, combined using
inverse-noise weighting, and Wiener filtered to suppress modes
which remain noisy in the combined set (as described in~\refsec{combine_qu}). 
We see that some modes remain unavailable for lensing template
construction at $|\ell_x|<100$ and $|\ell_y|>500$.
}
\label{fig:exptcomb}
\end{figure*}

\reffig{exptcomb} illustrates the process.
The left three panels show the 2D $E$-mode signal power spectra for
the three experiments.
We see the \lcdm\ $E$-mode spectrum rolled-off by the beam window
function of each telescope.
Because of their scan strategies and the applied scan-wise filtering, \bk\ and \sptpol\ 
have filtered out the modes along the $\ell_y$ axis;
while \planck\ has isotropic mode coverage.
The right panel shows the combined mode set after the final Wiener filter
step, so only modes measured with good signal-to-noise are retained.
At $\ell>500$ \planck\ does not have good per-mode signal-to-noise
so the modes along the $\ell_y$ axis beyond this multipole cannot
be filled in.

\begin{figure*}
\begin{center}
\includegraphics[width=\textwidth]{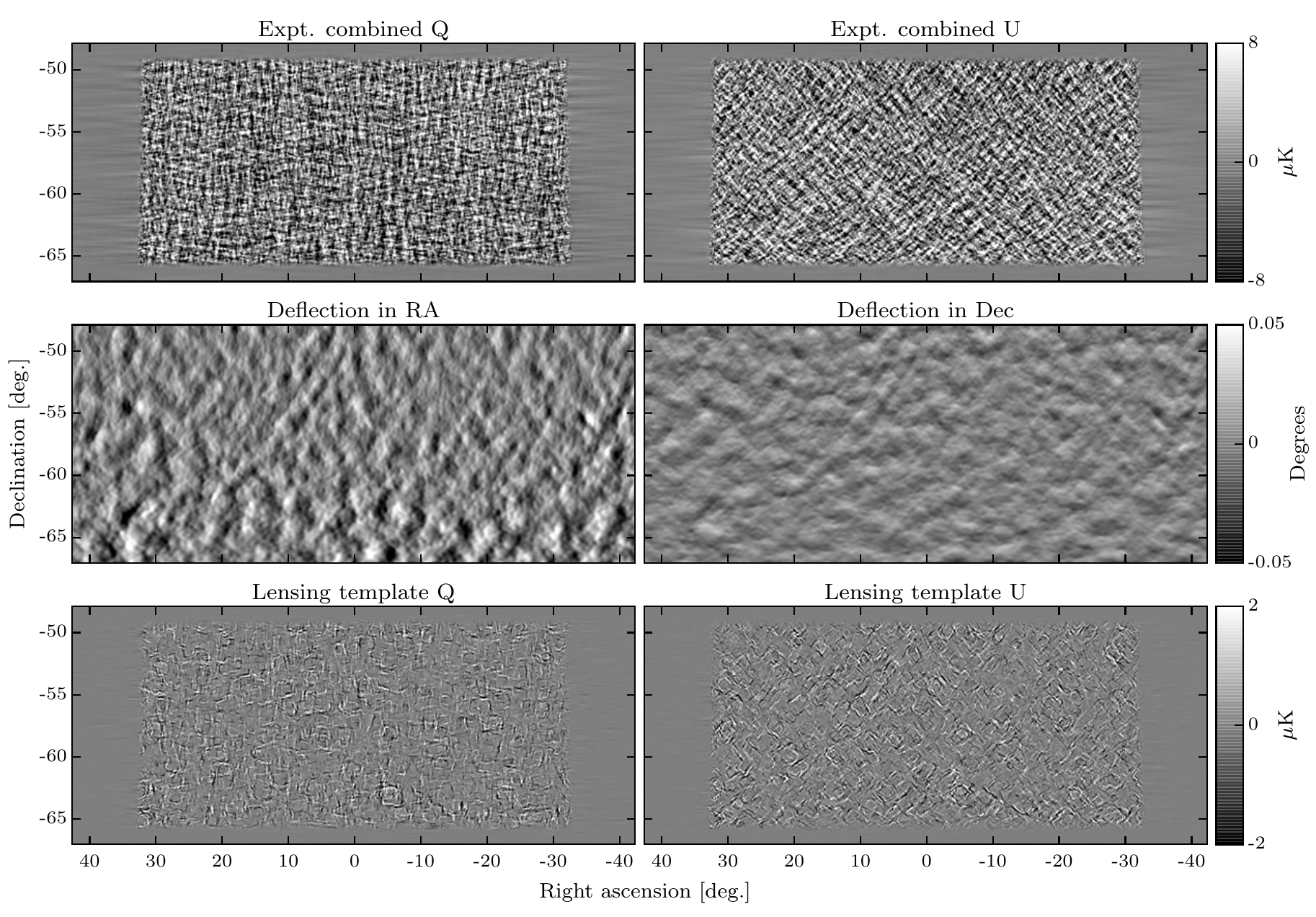}
\end{center}
\caption{The top two panels show the experiment-combined and Wiener filtered $Q/U$ maps.
  The middle two panels show the $x$ and $y$ derivatives of the normalized and
  Wiener filtered \planck\ CIB map.
  Signal and noise are approximately equal in these maps.
  Due to the foreshortening effect the RA deflections are larger and increase towards
  more negative Dec.
  The $Q/U$ maps in the top panels are undeflected by the angles shown in the middle panels
  and differenced with the initial maps to form the lensing template $Q/U$ maps shown in the bottom
  panels.
}
\label{fig:qu_defs_lt_maps}
\end{figure*}

We next proceed to inverse-Fourier transform the
combined and Wiener filtered $Q/U$ modes 
back to image space where they are ready to be undeflected by the gradient of the $\phi$ tracer.
The $Q/U$ maps at this stage are shown in the top panels of \reffig{qu_defs_lt_maps}.

\subsection{CIB map}\label{sec:cib_maps}
With the combined and filtered $Q/U$ maps in hand, we next need a $\phi$ tracer map.
In the following, we describe the characteristics of the CIB map used in this analysis,
and how we generate simulations of it in the \bicep/\keck\ patch.

\subsubsection{CIB data}\label{sec:cib_data}
It is possible to reconstruct the lensing potential field $\phi$ from
the CMB temperature and polarization patterns~\cite{plancklens18},
and in the future this will become the best $\phi$ estimate for delensing~\cite{smith2012}.
However, at the currently available noise levels the most effective available $\phi$
tracer is the CIB, even though it is only partially correlated with
$\phi$~\cite{simard15}.
Specifically, we use the 545~GHz CIB map from \planck\ generated using the
GNILC algorithm~\cite{planck2015cibdust}.\footnote{\href{http://pla.esac.esa.int/pla/aio/product-action?MAP.MAP_ID=COM_CompMap_CIB-GNILC-F545_2048_R2.00.fits}{CIB map: \texttt{COM\_CompMap\_CIB-GNILC-F545\_2048\_R2.00}}}
We also considered using the CIB maps generated by~\cite{lenz19} and will 
discuss that later in this section.
To determine the degree to which the GNILC CIB map is correlated with
$\phi$, we use the \planck\ 2015 minimum-variance lensing reconstruction
map\footnote{\href{https://irsa.ipac.caltech.edu/data/Planck/release_2/all-sky-maps/maps/component-maps/lensing/COM_CompMap_Lensing_2048_R2.00.tar}{\planck\ lensing map:  \texttt{COM\_CompMap\_Lensing\_2048\_R2.00}}}~\cite{plancklens15}
and make the assumption that this is an unbiased (although noisy) representation
of the true $\phi$ pattern.

We refer the reader to~\cite{2016A&A...596A.109P} for a detailed discussion of the \planck\ CIB map.
Briefly, the GNILC component-separation technique~\cite{Remazeilles11} disentangles different
components of emission using both frequency and spatial (angular-scale dependence) information.
In this case, the GNILC algorithm was applied to \planck\ data to disentangle Galactic dust emission and CIB anisotropies.
Even though both components share similar frequency spectral signatures, they have distinct angular power spectra.
Thus by using priors on the angular power spectra of the CIB, Galactic dust, the CMB,
and the instrumental noise, these components can be (partially) separated.
We note that the algorithm was developed mainly for extracting Galactic dust, 
and regions with different levels of Galactic dust can be expected to have 
different efficiencies of CIB recovery~\cite[e.g.][]{maniyar19, lenz19}. 
Therefore, in the following, we quantify the GNILC CIB map correlation 
with the \planck\ estimate of $\phi$ empirically in selected parts of the sky. 

\begin{figure}
\begin{center}
\includegraphics[width=0.48\textwidth]{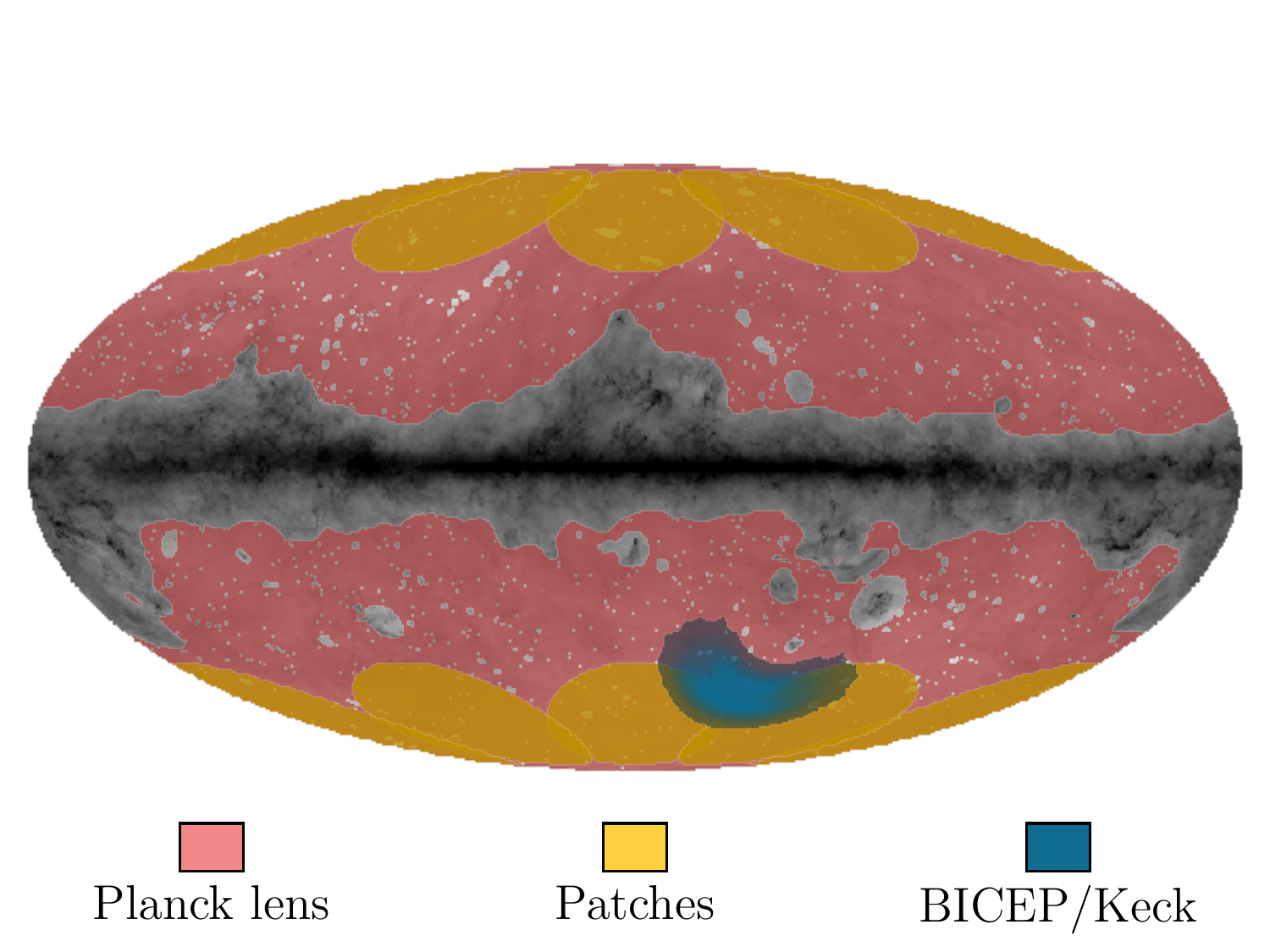}
\caption{
The light red regions denote the \planck\ lensing mask, used for computing the ``full sky" average of
GNILC CIB and $\phi'$ map cross-correlation. 
The yellow regions are the eight patches with similar size and unpolarized dust amplitudes
 as the \bk\ patch. These patches used for 
measuring the mean and scatter of the CIB auto-spectra and CIB$\times\phi'$, 
which are used as inputs to simulating CIB and filtering the CIB map. 
The overlaps between the yellow patches are small and apodization is applied when 
calculating the auto- and cross-spectra. 
The \bk\ patch is shaded in blue. 
The background is the \planck\ dust intensity map.}
\label{fig:patches}
\end{center}
\end{figure}

To select patches for estimating the CIB-$\phi$ correlations, 
we measure the mean amplitude in a \planck\ dust temperature 
map\footnote{\href{https://irsa.ipac.caltech.edu/data/Planck/release_2/all-sky-maps/previews/COM_CompMap_ThermalDust-commander_2048_R2.00/index.html}{Thermal dust emission map: \texttt{ThermalDust-commander\_2048\_R2.00}}}
of \mbox{$\sim500$ deg$^2$-sized} circles throughout the sky. 
Amongst the eight selected patches, as shown in~\reffig{patches}, 
the ratios of the mean amplitudes in the patches vs.~that in the 
\bk\ patch range from 0.6 to 1.7.  
These are thus similar to the \bk\ region in terms of their unpolarized dust intensities.

\begin{figure}
\begin{center}
\includegraphics[width=0.48\textwidth]{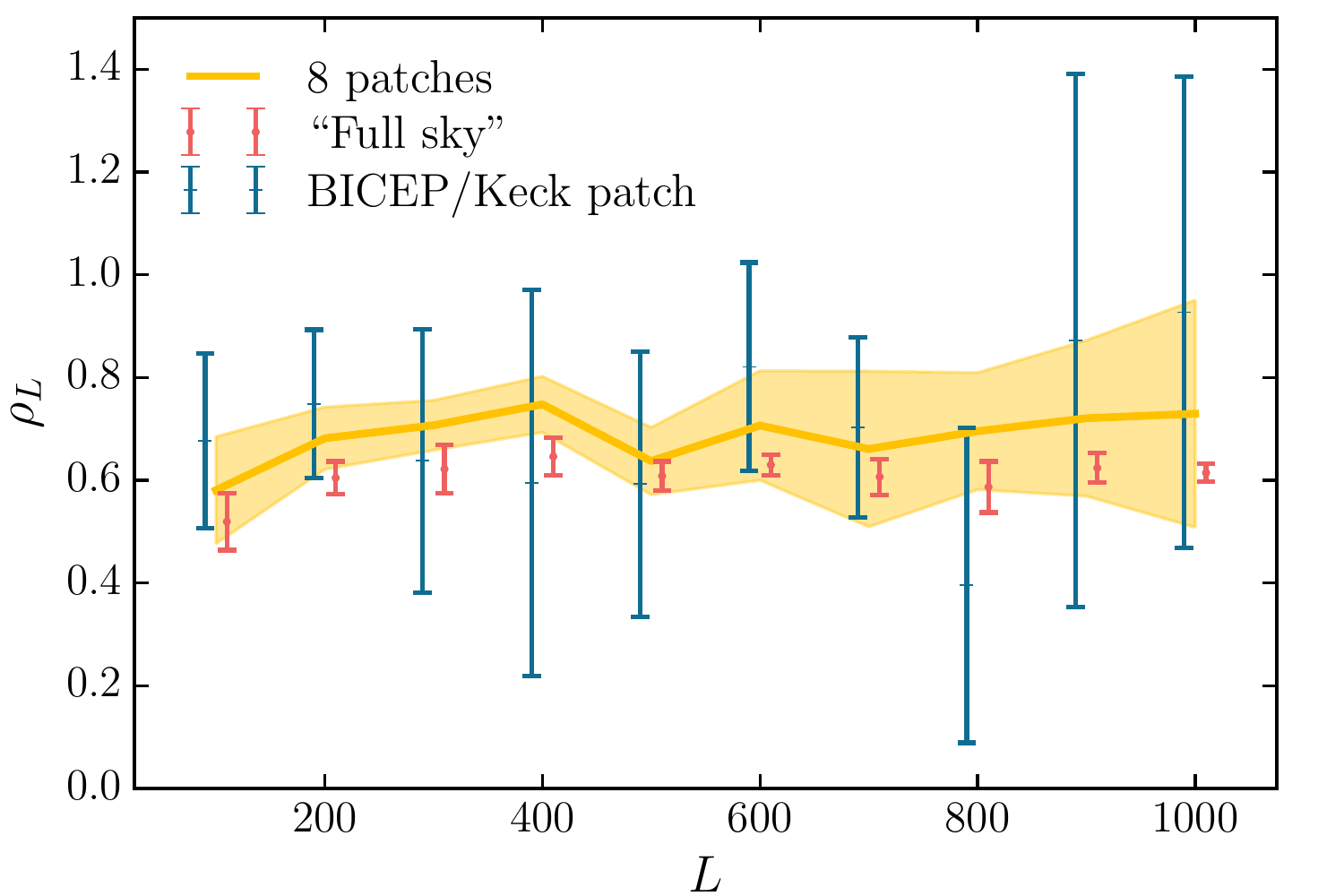}
\caption{The binned correlation factor $\rho_{L} = C_{L}^{\rm I\phi'}/\sqrt{C_{L}^{\rm II} C_{L}^{\phi\phi}}$ for the eight patches, the full sky, and the \bk\ patch. 
``Full sky'' corresponds to the overlap area between the \planck\ lensing mask and the GNILC CIB map. 
The $\rho_{L}$ in the \bk\ patch is consistent with those measured across the eight patches. 
The yellow band denoted by ``8 patches'' is the mean and standard deviation of $\rho_{L}$ across the 8 patches.
The error bars for the red and blue points are computed by taking the standard error of $\rho_{L}$ within 
each $\Delta L = 100$ bin. 
The red and blue points are shifted for clarity.
}
\label{fig:rhoell}
\end{center}
\end{figure}

In these patches, we compute the auto-spectra and cross-spectra using {\sc PolSpice}~\cite{chon04}.\footnote{\url{http://www2.iap.fr/users/hivon/software/PolSpice/README.html}}
We show the correlations $\rho_{L}$, defined as
$ C_{L}^{\rm I\phi'}/\sqrt{C_{L}^{\rm II} C_{L}^{\phi\phi}}$,
for a few different regions of the sky in~\reffig{rhoell}.
Here $I$ denotes the CIB map, $\phi'$ denotes the \planck\ lensing estimate, 
and $C_{L}^{\phi\phi}$ is the theory spectrum from the fiducial cosmology used in~\cite{planck2015cosmoparam}.
Comparing the correlations of the CIB map and the lensing map in the selected 
patches with that from the full overlap between the two maps (labeled ``Full Sky"), 
we observe that the correlations within the patches are higher than the 
correlation in the larger region that includes lower Galactic latitudes, and hence higher dust levels.
The full-overlap region correlation is $\sim 62\%$ for $L$ between 150 and 550, whereas
the mean correlation in the patches is $\sim 69\%$ over the same $L$ range.
\reffig{rhoell} also shows the cross-correlation in the \bk\ patch, which appears
to be consistent with the eight circular patches.

As a cross check, 
we compare within the \bk\ patch the cross-spectrum of the GNILC CIB map and the \planck\ lensing map 
against the cross-spectrum of 
a CIB map produced by \cite{lenz19} and the \planck\ lensing map. This CIB map has been cleaned 
using neutral hydrogen (HI) as a Galactic foreground tracer, with an HI column density threshold of $2.5\times10^{20}$\, cm$^{-2}$.
We find the lensing correlation in the two CIB maps to be consistent with each other,
thus providing additional evidence that in the \bk\ map region, 
the GNILC CIB map does not show the reduced correlation which is expected,
and seen, in regions closer to the Galactic plane.

The filter and normalization of the $\phi$ tracer is given in general form in \refeqn{tracer_wiener}. 
In this case, we take it as the average over the eight patches of the cross-spectra of the CIB and the
lensing map divided by the average of the CIB auto-spectra:
\beq
\phi^{\rm CIB}_{LM}  = \left( \frac{ \langle C_{L}^{\rm I\phi'} \rangle_{\rm patches}}{ \langle C_{L}^{\rm II} \rangle_{\rm patches}} \right) I_{LM}.
\eeq
To further prevent Galactic dust contamination, we additionally impose a $L_{\rm min} = 100$ cut.
This filter and normalization is applied to the real data as well as the simulated CIB realizations which
are described in the next section.
We render the normalized and Wiener filtered CIB $a_{\ell m}$ and its associated gradients to \healpix\ maps
of $N_{\rm side} =512$, 
and then interpolate and convert the gradient maps to derivatives with respect to our pixel grid. 
The derivatives are shown in the middle panels of \reffig{qu_defs_lt_maps}.

\subsubsection{CIB simulations}
\label{sec:cib_sims}

We use CIB simulations to estimate the expected level of 
lensing $B$-modes in the lensing template, to form the bandpower covariance in the 
likelihood analysis, and as inputs to null tests. 

We generate CIB simulations based on the input Gaussian $\phi$ fields of the \bk\ 
simulation set described in~\refsec{cmbsims}. 
To convert the $\phi$ fields to CIB fields, we use the auto-spectrum
of the CIB, \cii, and the cross-spectrum of the CIB and the \planck\
lensing estimate $\phi'$, \ciphi.
We construct each CIB field by rescaling each input $\phi$ field 
so that the cross-spectrum of the rescaled field with the input $\phi$ is \ciphi. 
We then add to the rescaled $\phi$ field Gaussian noise so that its auto-spectrum is \cii. 
Formally, we construct the signal part of the CIB simulations, $I_{LM}^S$, as
\beq
I_{LM}^S = \frac{ C_{L}^{\rm I\phi'}}{C_{L}^{\phi\phi}} \phi_{LM}
\label{eqn:cib_gauss}
\eeq
where $\phi_{LM}$ are the spherical harmonic coefficients of the input $\phi$ fields.
We construct the noise part of the CIB simulations,  $I_{LM}^N$, by
generating Gaussian random fields with power spectrum described by
\cii $-$ (\ciphi)$^2/ C_{L}^{\phi\phi}$. 
The total CIB field is the sum of the two terms
$I_{LM}  =  I_{LM}^S + I_{LM}^N$. 

We have 499 realizations of $\phi_{LM}$. 
For each $\phi_{LM}$, we form $I_{LM}$ as described in the previous paragraph 
with $C_{L}^{\phi\phi}$ from the input theory\footnote{Here we have taken $C_{L}^{\phi\phi}$
as the \planck\ 2013 cosmology used to generate the \bk\ simulations introduced
in~\refsec{cmbsims}.
This is slightly different than the latest \planck\ cosmology which is implicit in
the $C_{L}^{\rm I\phi'}$ of~\refeqn{cib_gauss}.
Arguably it would be more self-consistent to use the latest $C_{L}^{\phi\phi}$ here.
However, we have checked that this makes no practical difference at
the current sensitivity level.}
and \cii\ and \ciphi\ sampled from the measured mean
and covariance of \cii\ and \ciphi\ from the 8 patches selected in~\refsec{cib_data}.
In the limit of many realizations, the simulated $I_{LM}$ will have the same covariance structure
in \cii\ and \ciphi\ as measured from the 8 patches.
The advantage of sampling \cii\ and \ciphi\ as opposed to using the measured mean from the 8 patches
is that the potential patch-to-patch variation of the CIB auto-spectrum, 
and the cross-spectrum between CIB and $\phi'$, is built into the simulations.
Therefore, the uncertainties in the CIB measurements are propagated to the
uncertainty in the $r$ measurement.

At this point we use the method described
in~\refsec{construct_lt} to undeflect the combined $Q/U$ data and simulation maps
with the data and simulation CIB maps to form the real and simulated lensing templates.
The lensing templates from the real data are shown in the bottom panels of \reffig{qu_defs_lt_maps}.

We have now laid out the lensing template construction, the extension of the \bk\ 
analysis framework, and the input simulations and data used in this paper. 
The next steps include demonstrating the robustness of these extensions
to potential biases and misestimations of inputs.

\section{Pipeline and Simulation Validation}
\label{sec:simval}

In this section, we demonstrate the robustness of the pipeline in the limit of perfect delensing,
quantify the level of bias to our inference of $r$ 
given potential misestimations in the inputs to our simulations, and
estimate the impact on $\sigma(r)$ given variations in the simulation setup.
To do that we use the set of lensed-\lcdm+dust+noise simulations ($r=0$)
from the BK14 paper.
We run maximum-likelihood searches of the baseline lensed-\lcdm+dust+synchrotron+$r$
model as described in Appendix E.3 of the BK14 paper, in this case
adding a lensing template.

\subsection{$r$ recovery with perfect delensing}
To validate the addition of the lensing template as a pseudo-band in the \bk\ analysis framework,
we run maximum-likelihood searches in two configurations---unlensed input CMB skies
without lensing templates,
and lensed input CMB skies with perfect lensing templates.
The perfect lensing templates are constructed by differencing the filtered, noiseless, lensed and
unlensed $Q/U$ skies.
If the likelihood works as intended, 
we expect the recovered $r$ values from the two sets of simulations
to be extremely close to each other on a realization-by-realization basis.
We find that the differences between the recovered $r$ values
$|\Delta r| \lesssim 0.002$. 
We also find that at our current noise level, even if we have perfect knowledge of the 
lensing $B$-modes in our patch,
the uncertainty on $r$ is reduced only by 26\% from $\sigma(r) = \simbksr$ to $\sigma(r) = \simbkperfectltsr$. 
This means that lensing uncertainty is subdominant compared to uncertainties from
foregrounds and instrument noise in the BK14 data set.

\subsection{Biases to $r$ from misestimations of inputs}

We investigate the bias to $r$ from the following:
(1) misestimation of the correlation between the CIB map and $\phi$, and
(2) biases in polarization efficiency in the input $Q/U$ maps.

\textit{Misestimation of $C_{L}^{\text{I}\phi}$ :}
As discussed in~\refsec{cib_data}, we compute \cii\ and \ciphi\ used in the normalization and Wiener
filter of the CIB map as the mean of the \cii\ and \ciphi\ from 8 patches. 
In addition, we generate simulations of CIB based on the mean and scatter 
of the \cii\ and \ciphi\ spectra measured from the patches.
Here, we consider the case in which the actual CIB cross-spectrum
with $\phi$ is offset from the measured mean \ciphi. 
A plausible way in which the measured \ciphi\ might be biased from the true $C_{L}^{\text{I}\phi}$ is through
a bias in the $\phi'$ reconstruction due to CIB in the input temperature maps.
In that case, the measured \ciphi\ would contain a term that comes from 
CIB$\times \phi( {\rm CIB, CIB})$, where $\phi( {\rm CIB, CIB})$ denotes the 
CIB power that is leaked through the $\phi$ estimator applied to the CMB maps.

We construct a test for this bias, which proceeds as follows:
using the measured mean \cii\ and \ciphi, we generate simulated CIB skies as described in~\refsec{cib_sims}.
This set of simulations is the assumed truth.
We then generate 2 sets of CIB skies whose \ciphi\ is either half a $\sigma_{\rm sp}$ above or below the measured mean \ciphi, 
where $\sigma_{\rm sp}$ is measured from the spread across the 8 patches.
We process these CIB skies as if they had the mean \ciphi, i.e. 
we normalize and Wiener filter these maps using the mean \ciphi\ and \cii.
We then proceed to construct lensing templates and
calculate auto- and cross-spectra with the rest of the BK14 maps,
exactly as in the baseline analysis.
The bandpower covariance matrix is derived using lensing templates constructed from 
the nominal, unbiased set of CIB skies.
We then run maximum-likelihood searches on these two sets of simulations for the
model parameters $r$, $\Ad$, $\As$, $\Bd$, and $\Bs$.
We determine the bias on $r$ by comparing the means of the
maximum likelihood $r$ values from the nominal set and the half-$\sigma_{\rm sp}$ offset sets.
We observe that the mean $r$ is biased by 0.2$\sigma$, where $\sigma$ denotes
the uncertainty of the $r$ measurement (i.e.\ the width of the $r$ distribution of the nominal set). 

To get a sense of how relevant this bias is, we compare the
half-$\sigma_{\rm sp}$ offset we introduce into the simulations to a worst-case scenario of 
CIB leakage in the reconstructed $\phi'$ map.
Reference \cite{omori17} estimated the term CIB$\times \phi({\rm CIB, CIB})$ using 
$\phi'$ reconstructed from the \planck\ 545~GHz maps without 
foreground cleaning and found the bias to be below $\sim$5\% for $L < 1024$, the $L$ 
range used in this work. 
A 5\% bias is smaller than the half $\sigma_{\rm sp}$ shift considered.
Furthermore, the \planck\ lensing map used to calculate \ciphi\ was constructed using 
the SMICA input maps that are foreground-suppressed. 
Therefore, we expect the 0.2$\sigma$ bias to be an overestimate of 
potential biases from misestimating \ciphi. 

\textit{Misestimation of polarization efficiency:}
The \planck\ collaboration has found that their polarization efficiency calibration could
potentially be biased at the 1--2\% level~\cite[see e.g. Table 9 of][]{planck2018III}. 
The \sptpol\ $Q/U$ maps are calibrated using a \planck\ $E$-mode map~\cite{henning18}.
Therefore, it is reasonable to ask by how much $r$ would be biased if the calibration of
the input $Q/U$ maps is biased.

We construct the test by artificially scaling the \sptpol\ $Q/U$ simulated maps low by 1.7\%
and analyzing the maps as if they had the original amplitudes. 
In other words, similar to the half-$\sigma_{\rm sp}$ \ciphi\ shift test above, the rest of the 
pipeline is held identical and the only change is the input \sptpol\ $Q/U$ maps.
For simplicity, instead of using the combined $Q/U$ map, we use only \sptpol\ simulated
maps for this test. 
Comparing the mean of the recovered maximum likelihood $r$ values
for the nominal set with that for the biased set, we find negligible differences.
We conclude that biases at this level in the polarization efficiency are not an issue for
this analysis. 

\subsection{Impact on $\sigma(r)$ from variations of inputs}

We investigate the impact on $\sigma(r)$ from two effects:
(1) non-Gaussianities in the input CIB map, and
(2) inclusion of patch-to-patch variation in \cii\ and \ciphi in the generation
of the CIB realizations.

\textit{Non-Gaussianity of the CIB:}
As discussed in~\refsec{cib_sims}, we generate our CIB simulations based on the $\phi$
realizations used to lens the simulated CMB input skies. 
While the $\phi$ realizations are Gaussian, 
the true $\phi$ has some non-Gaussianities due to non-linear growth of structure~\cite{boehm16}. 
However, the contribution to lensing $B$-modes from non-Gaussian $\phi$ is subdominant
over the angular scales considered~\cite{lewis2016}. 
It is thus sufficient to model $\phi$ and the portion of CIB that correlates with $\phi$, the
signal term (\refeqn{cib_gauss}), as Gaussian. 
In addition to the signal term, we simulate the noise term of the CIB $I_{LM}^N$---the portion 
of the CIB that does not correlate with $\phi$---as Gaussian realizations given 
the measured \cii, \ciphi, and the input $C_{L}^{\phi\phi}$.
However, the CIB is known to be quite non-Gaussian; its bispectra at the angular scales relevant 
to this work have been measured by~\cite{planck2013XXX} with high signal-to-noise.
Therefore, one could imagine that simulating the CIB $I_{LM}^N$ as Gaussian fluctuations 
would cause the lensing template fluctuation to be underestimated.
With the underestimation of the lensing template fluctuation, 
$\sigma(r)$ would be underestimated. 
Here we estimate the impact on $\sigma(r)$ when we increase the lensing template 
fluctuation. 

To get a handle on how much to increase the lensing template fluctuation, 
we build lensing templates using a simulated CIB sky from Websky 
mocks,\footnote{\url{https://mocks.cita.utoronto.ca/index.php/WebSky\_Extragalactic\_CMB\_Mocks}}
which are built based on an approximation to full N-body halo catalogs~\cite{stein18, stein20}.
From the full-sky CIB realization, we
make 80 cutouts of size similar to the \bk\ patch, undeflect-and-difference the $Q/U$ maps, 
 and compute the lensing template bandpower variances. 
We then generate matching Gaussian realizations of CIB using the \cii\ and $C_{L}^{\text{I}\phi}$
measured between the simulated CIB map and the corresponding $\phi$ map 
(provided as a $\kappa$ map, where $\kappa = - \nabla^2 \phi/2$).
Using these Gaussian CIB realizations, we generate lensing templates and calculate their
bandpower variances. 
For the $L$ range considered in this analysis, the ratio of the lensing template 1$\sigma$
uncertainties between templates generated from Gaussian CIB and those from N-body based 
CIB is 0.97 $\pm$ 0.07. 
This suggests that the lensing template bandpower variance is sufficiently modeled using
Gaussian simulations. 
Furthermore, \cite{namikawa19} performed a similar test using galaxy densities as $\phi$ tracer
and  found that the difference in the lensing template covariance 
between the Gaussian and their simulations is within the Monte-Carlo uncertainty of the number of
simulations considered.

Since the above tests could still be limited by the number of non-Gaussian simulated CIB skies, 
we ask how much $\sigma(r)$ could be impacted because of some low
level of unmodeled non-Gaussianity in the $\phi$ tracer. 
To do that, we increase the values in the lensing template auto-spectrum sub-block 
of the bandpower covariance matrix by 10\% and perform maximum-likelihood searches
on the baseline set of simulations. 
The resultant $\sigma(r)$ estimated from the width of the $r$ value distribution
is negligibly different to the baseline case. 
Therefore, we conclude that at the current level of noise, unmodeled non-Gaussianities of the CIB 
have negligible impact on the uncertainty of the $r$ measurement.

\textit{Patch-to-patch variation in \cii\ and \ciphi:}
We construct the CIB realizations using samples of \cii\ and \ciphi\ drawn 
from the measured covariance of \cii\ and \ciphi\ across 8 patches.
By doing this we incorporate the patch-to-patch variation in the CIB auto- and cross-spectrum
with $\phi'$ into the uncertainty on $r$.
Here we check how large this effect is by comparing the $\sigma(r)$ estimated
from a set of CIB simulations generated with fixed \cii\ and \ciphi\ with that 
estimated from a set of CIB simulations generated from a distribution of \cii\ and \ciphi. 
We find that the $\sigma(r)$ from these two sets of simulations are compatible
to within MC uncertainty.
This means that the uncertainty on $r$ introduced by the uncertainties in \cii\ and \ciphi\ 
is subdominant compared with the noise and sample variance of the lensing templates.

Having estimated the biases to $r$ caused by possible biases in the CIB and $Q/U$ maps 
and found them to be small, 
and having shown the impact on $\sigma(r)$ due 
to unmodeled non-Gaussianity of the CIB to be minimal, we now turn to testing
the robustness of the simulations against unmodeled Galactic foregrounds using the data themselves.

\section{Systematics checks}\label{sec:syscheck}
\label{sec:sys}
Previous \bk\ papers include ``jackknife'' internal consistency tests
on the 95 and 150\,GHz maps used here~\cite{bk1,bk5,bk6}.
In this section, we provide similar tests of the auto- and cross-spectra
of the newly introduced lensing template.
We consider the following ways in which
the simulations can fail to sufficiently describe the statistics of the
lensing template:
\begin{enumerate}
\item Galactic dust in the input 150~GHz $Q/U$ maps leaks into the lensing template,
\item low-$\ell$ systematic residuals in the \planck\ polarization maps leak into the lensing template, 
\item non-Gaussian Galactic dust residuals in the CIB map introduce extra power in the lensing
template beyond that described by Gaussian modeling of uncorrelated power.
\end{enumerate}
All of the above would (i) increase the power of the lensing template auto-spectrum,
and (ii) introduce potential chance coupling with the observed $B$-modes.

Galactic dust power is sub-dominant to $E$-mode power over the angular scales
relevant to producing the lensing $B$-mode template, and
the simulated $Q/U$ maps used in \refsec{cmbsims} do not include a dust component.
However, we would still like to check that
the Galactic dust component in the $Q/U$ data maps does not significantly contribute to the lensing 
template auto-spectrum.
For the CIB map, any components that contribute to the CIB auto-spectrum but
are uncorrelated with $\phi'$ are modeled as Gaussian fluctuations. 
Therefore, the unmodeled non-Gaussian Galactic foregrounds could contribute extra fluctuation
in the lensing templates when used to undeflect the CMB maps.
In addition, they could contribute extra template power when deflecting the unmodeled
Galactic foregrounds in the $Q/U$ maps.

\begin{figure*}
\begin{center}
\includegraphics[width=0.98\textwidth]{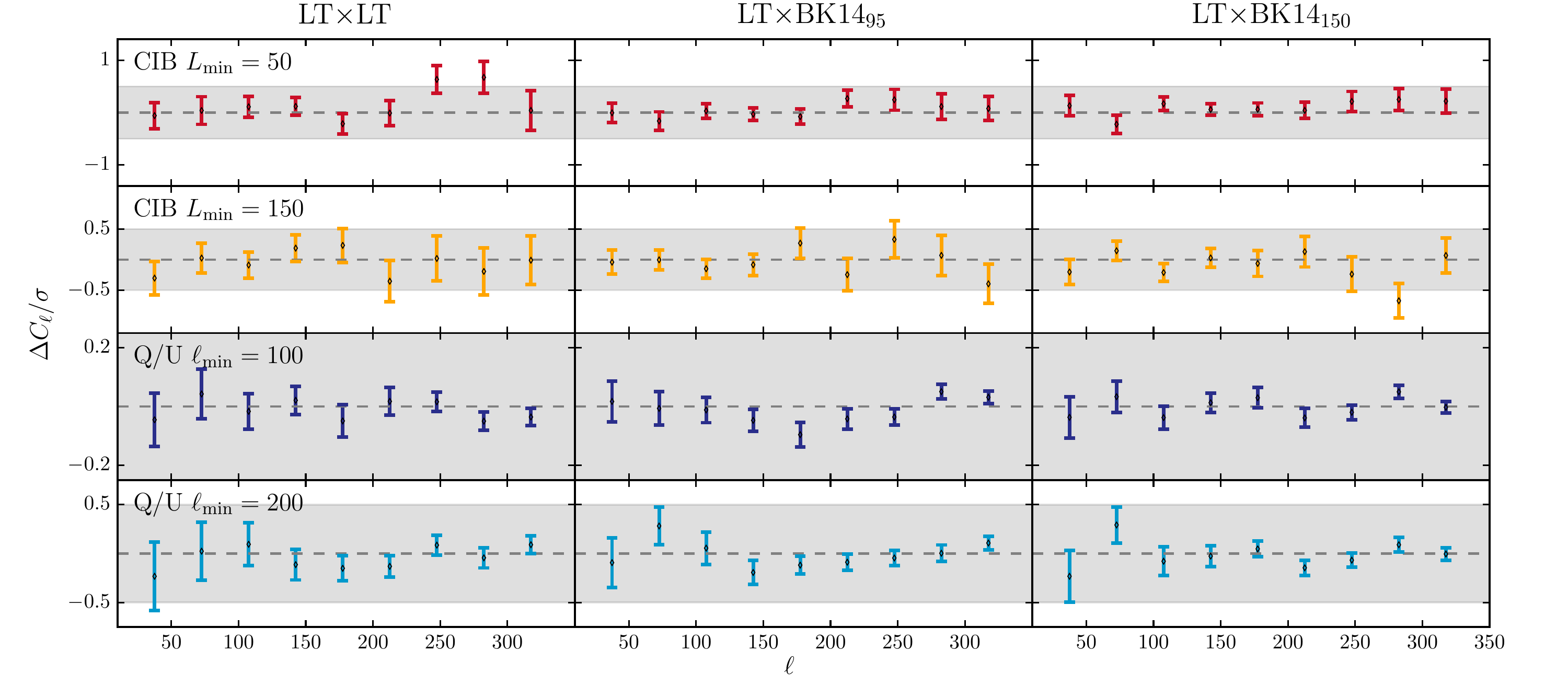}
\caption{
Difference bandpowers ($\Delta C_{\ell}$, see definition in text) between the baseline analysis and 
analyses with one parameter changed, and the uncertainties on those difference bandpowers, 
both scaled by the statistical uncertainties on the baseline analysis bandpowers. 
The label at the top left hand corner of each row indicates 
which parameter has been modified and how it is modified. 
The left to right columns show the difference bandpowers from the lensing template auto-spectrum, 
lensing template cross-spectrum with the BK14 95~GHz map, 
and lensing template cross-spectrum with the BK14 150~GHz map.
The gray bands indicate the 0.5$\sigma$ statistical uncertainty 
of the baseline spectra. 
The $\chi_{\rm sys}^2$ and PTE of the difference bandpowers are listed in~\reftab{chi2}.
We find the data difference bandpowers to be consistent with the spread in the simulation difference bandpowers.
}
\label{fig:deltasigma}
\end{center}
\end{figure*}

To address the question of whether the simulations are a sufficient description of the data
given these unmodeled effects, we test the consistency of the lensing template auto- and cross-spectra
against simulations.
Specifically, we perform spectrum-difference tests where we compare the
difference spectrum of data between the baseline $\ell$ \& $L$ ranges and
variant $\ell$ \& $L$ ranges against the corresponding differences in simulation.
We calculate two quantities, $\chi^2$ and $\chi$, as follows.
Firstly
\beq
\chi^2_{\rm sys} =   \Delta C_{\ell}^{\dagger}\, {\rm Cov^{-1}} \Delta C_{\ell},
\eeq
where $\Delta C_{\ell}$ denotes the binned data difference spectrum and ${\rm Cov}$ is the bandpower covariance
matrix formed from the difference spectra of the corresponding simulations.
And secondly
\beq
\chi_{\rm sys} = \sum_{\ell} \Delta C_{\ell} / \sigma_{\ell, \rm diff},
\eeq
where $\sigma_{\ell, \rm diff}$ denotes the standard deviation from the simulation difference spectra.
\reffig{deltasigma} shows the difference spectra for the lensing template auto-spectrum (LT$\times$LT), 
lensing template cross-spectrum with the BK14 95~GHz map (LT$\times$BK14$_{\rm 95}$), and
lensing template cross-spectrum with the BK14 150~GHz map (LT$\times$BK14$_{\rm 150}$). 
The PTE values from $\chi^2_{\rm sys}$ and $\chi_{\rm sys}$ are listed in~\reftab{chi2}.

\begin{table}
\caption{The PTE values from $\chi^2_{\rm sys}$ and $\chi_{\rm sys}$ 
(separated by a comma)
with different CIB input $L_{\rm min}$ 
and $Q/U$ map input $\ell_{\rm min}$, compared with the baseline setup.
LT$\times$LT , LT$\times$95, and LT$\times$150 denote the lensing template
auto-spectrum, lensing template cross-spectrum with the BK14 95~GHz map, and 
with the BK14 150~GHz map, respectively.}
\centering
\begin{tabular}{l |r |r| r }
\hline\hline
Variation /\ spectrum & LT$\times$LT &  LT$\times$95  &  LT$\times$150    \\
\hline
CIB $L_{\rm min}$ = 50     & 0.36, 0.12  & 0.80, 0.23 & 0.66, 0.09    \\
CIB $L_{\rm min}$ = 150    &  0.91, 0.67  & 0.68, 0.63 &   0.25, 0.88   \\
$Q/U$ $\ell_{\rm min}$ = 100  & 0.76, 0.70  &  0.09, 0.84  & 0.34, 0.52 \\
$Q/U$ $\ell_{\rm min}$ = 200   & 0.76, 0.75  & 0.36, 0.57 &  0.28, 0.62 \\
\hline
\end{tabular}
\vspace*{2mm}
\label{tab:chi2}
\vspace{0.1cm}
\end{table}

\subsection{\texorpdfstring{$L$}{L}-cuts on CIB map}
At large angular scales the CIB map could be contaminated by Galactic dust 
and thus a test in which the $L_{\rm min}$ for the CIB map is varied could be sensitive to its impact.
The unmodeled non-Gaussianity of residual Galactic foregrounds in 
the CIB map would cause the lensing template to have larger variance
than it would otherwise. 
We test the hypothesis that
the simulations are sufficient descriptions of the real data
by differencing the lensing template auto- and cross-spectra 
generated using the baseline $L_{\rm min} = 100$ for the CIB map and those
generated with $L_{\rm min} = 50$ and $L_{\rm min} = 150$. 
The PTEs from the $L_{\rm min}$ difference spectra show
that the data differences are sufficiently described by 
the simulation-difference distributions. 

 \subsection{\texorpdfstring{$\ell$}{L}-cuts on $Q/U$ maps}
Galactic dust contributes a fraction of the total power in
the $Q/U$ maps on the largest scales. 
Additionally, there could be low levels of unmodeled systematic residuals~\cite{planck2016SRoll} in the \planck\ $Q/U$ maps
that could leak power to the lensing templates. 
Similar to the test done with the CIB map, we set the $\ell_{\rm min}$ of the input $Q/U$ map
to two different levels compared to the baseline (no explicit $\ell_{\rm min}$ set)
and compute difference spectra between the variant $\ell_{\rm min}$ and the baseline.
For $\ell_{\rm min} = 100$ and $\ell_{\rm min} = 200$, we find the PTEs
from the difference spectra to be consistent with the simulation-difference distributions.

We thus conclude that at the current level of noise, the lensing template auto-spectrum and the
cross-spectra with the 95~GHz and 150~GHz maps do not contain unmodeled systematics
from large angular scales of the input $Q/U$ and CIB maps large enough to be incompatible with the
simulation distributions.

\section{Results}\label{sec:results}
We now proceed to repeat the parameter constraint analysis from the BK14 paper~\cite{bk6}
including the lensing template extension described and validated above.
We present two main results in this work.
First, we estimate $\sigma(r)$ with delensing by running maximum-likelihood
searches on the set of lensed-\lcdm+dust+noise simulations from BK14.
Second, we explore the likelihood space of the real data
and provide constraints on $r$ and the foreground model parameters.

\begin{figure*}
\begin{center}
\includegraphics[width=\textwidth]{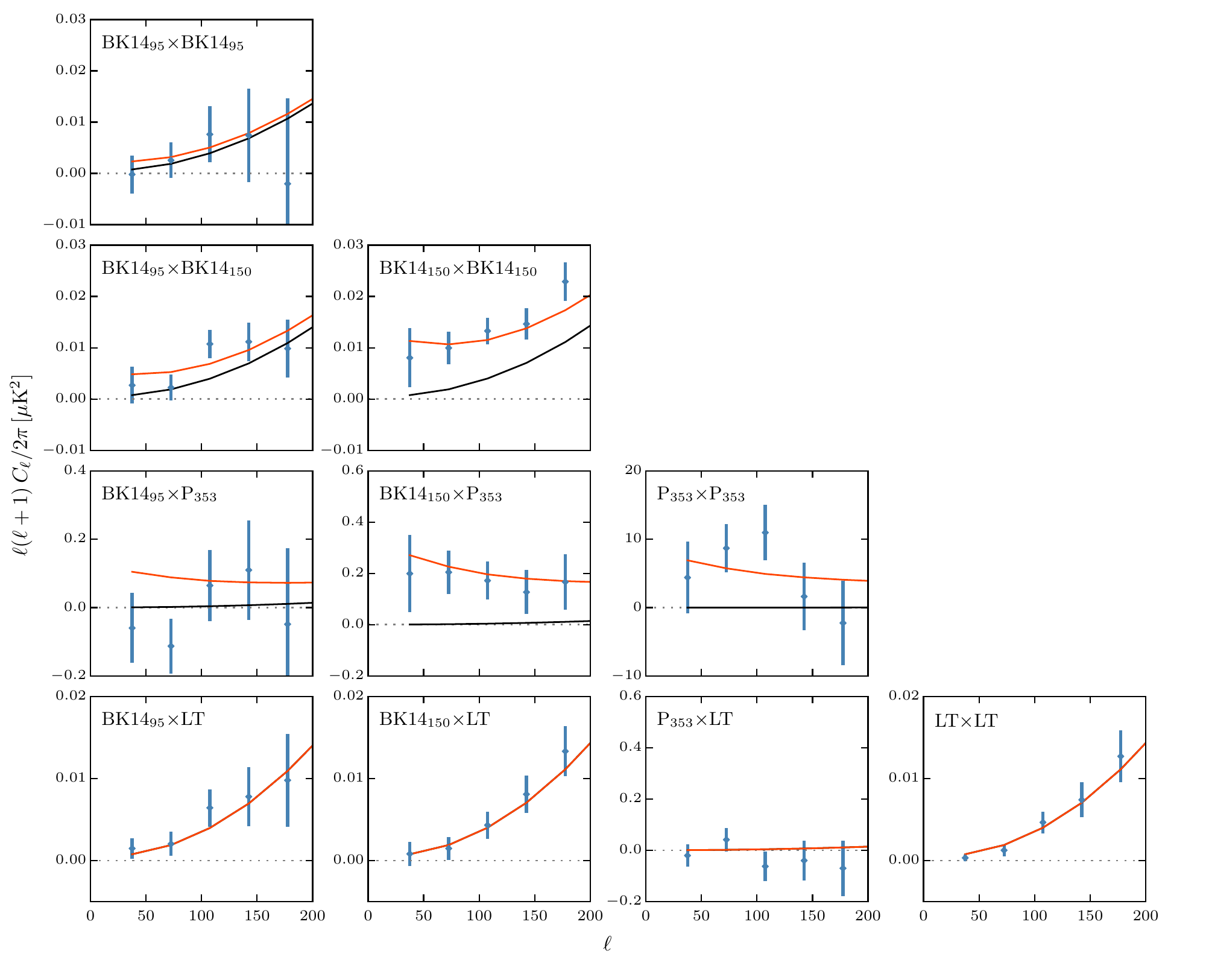}
\caption{$BB$ auto- and cross-spectra calculated using
\biceptwo/\keck\ 95 \& 150\,GHz maps, the \planck\ 353\,GHz
map, and the lensing template developed in this paper.
The black lines show the model expectation values
for lensed-\lcdm, while the red lines show the expectation values of the
baseline lensed-\lcdm+dust model from the BK14 analysis
($r=0$, $\Ad=4.3$\,\uksq, $\Bd=1.6$, $\ad=-0.4$),
and the error bars are scaled to that model.
Compared to the BK14 $BB$ spectra, which contain both foregrounds
and lensing components, 
the lensing template represents an alternate way to estimate
the lensing $B$-modes which is largely foreground-immune, and,
as we see here, provides good signal-to-noise in the resulting auto-
and cross-spectra.
}
\label{fig:bandpowers}
\end{center}
\end{figure*}

In \refsec{datasims}, we described the construction of a lensing
template using the \planck\ GNILC CIB map
and the combined $Q/U$ maps from \sptpol, \bk, and \planck.
\reffig{bandpowers} shows the auto- and cross-spectra of this
lensing template with the maps that most significantly constrain 
the model parameters---the \bk\ 95 \& 150\,GHz maps, and the
\planck\ 353\,GHz map.
The lensing template auto- and cross-spectra shown in~\reffig{bandpowers}, 
plus the additional cross-spectra 
with the other bands of \wmap\ and \planck, are the new
additions to the bandpower data vector input
to the likelihood analysis.
It is interesting to note that the error bars are much smaller at low $\ell$ for
LT$\times$LT than for LT$\times$150.
This is because, although the 150\,GHz map noise is very small, the
dust sample variance is large.

\subsection{Reduction in \texorpdfstring{$\sigma(r)$}{sigma r}}
\label{sec:sim_sr}
The inclusion of the lensing template cross-spectra reduces the effective sample variance
of the lensing component of the observed $B$-modes.
This is the reason that the uncertainty of the $r$ component can be 
reduced when we add a lensing template to the likelihood.

\begin{figure*}
\begin{center}
\includegraphics[width=\textwidth]{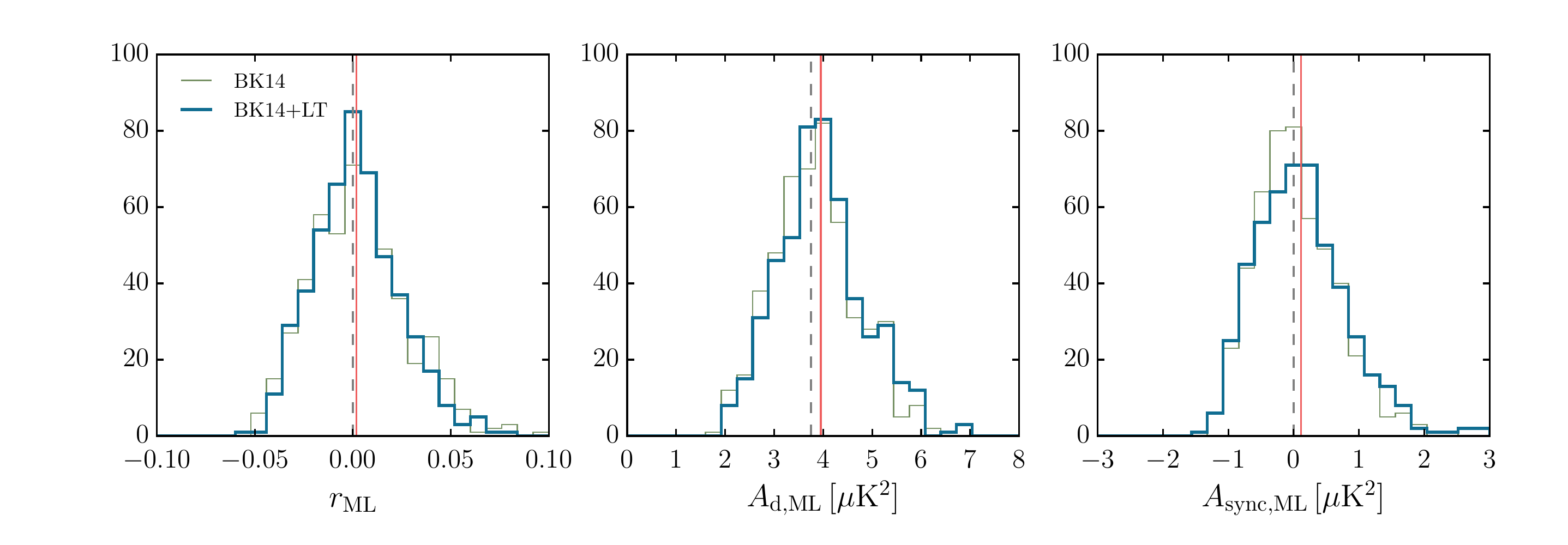}
\caption{Histograms of maximum-likelihood values of $r$, $\Ad$, and $\As$
from 499 realizations of BK14+LT (blue) and BK14 (gray) lensed-\lcdm+dust+noise simulations
in the baseline model with 6 free parameters: $r$, $\Ad$, $\As$,
$\Bd$, $\Bs$ and $\ad$.
The red lines mark the means of the distributions for the BK14+LT simulation set 
and the gray dashed lines mark the input values.
$\sigma(r)$ from the BK14+LT (BK14) simulation set is \simbkltsr\ (\simbksr) from the leftmost panel.
}
\label{fig:mlhist}
\end{center}
\end{figure*}

In BK14, we introduced $\sigma(r)$ as a measure of the intrinsic constraining power
of a given set of experimental data.
In contrast to the width of the 68\% highest posterior density interval as derived from
the real data this measure is not subject to noise fluctuation within that single
realization.
To compare the $\sigma(r)$ from the BK14 data set and the BK14 data set
with lensing template included, 
we repeat the analysis of Appendix E.3 of the BK14 paper.
We run maximum-likelihood searches with the baseline lensed-\lcdm+dust+synchrotron+$r$ model
on the lensed-\lcdm+dust+noise simulations for the two cases.
The parameters and priors are the same as in BK14 and are summarized in \reftab{priors}.
The amplitudes at $\ell=80$ of the dust and synchrotron $BB$ spectra
defined at 353~GHz and 23~GHz are denoted by $\Ad$ and $\As$, respectively; 
$\beta$ and $\alpha$ denote the frequency and spatial spectral indices,
with subscripts $d$ and $s$ referring to dust and synchrotron respectively;
$\epsilon$ denotes the dust-synchrotron correlation.
Flat priors are applied to $r$, $\Ad$, $\As$
\& $\ad$, and Gaussian priors are applied to $\Bd$ \& $\Bs$.
\reffig{mlhist} shows the distributions of maximum likelihood $r$, 
$\Ad$, and $\As$ values.
With the inclusion of the lensing template, we reduce $\sigma(r)$
from \simbksr\ to \simbkltsr, 
a $\sim10\%$ reduction.\footnote{Note that this $\sigma(r)$ is computed in a 6-dimensional parameter
space as opposed to the 8-dimensional parameter space which is used when sampling.
This is to maintain consistency with the BK14 paper. 
For a 8D search, $\sigma(r) = 0.026$ without the lensing template and $\sigma(r) = 0.023$ with.
The relevant metric here is the fractional reduction in $\sigma(r)$ between the two simulation sets 
which is similar for the 6D and 8D searches.
} 

\begin{table}
\caption{Priors imposed on each parameter for both maximum-likelihood search 
and posterior sampling for the baseline analysis.
$\mathcal{U}(a,b)$ denotes uniform distribution between $[a,b]$. 
$\mathcal{N}(\mu, \sigma^2)$ denotes normal distribution with mean $\mu$
and variance $\sigma^2$.
}
\centering
\begin{tabular}{ C{2cm} | C{3cm} | C{3cm}}
\hline\hline
Parameter & ML search & Sampling    \\
\hline
 $r$      &  $\mathcal{U} (-0.5,0.5) $  & $\mathcal{U} (0, 0.5) $\\
$\Ad$     &  $\mathcal{U} (-2, 15) $ & $\mathcal{U} (0, 15) $ \\
$\As$     &  $\mathcal{U} (-2, 15) $ & $\mathcal{U} (0, 50) $  \\
$\Bd$     & $\mathcal{N} (1.6, 0.11^2) $ & $\mathcal{N} (1.59, 0.11^2) $   \\
$\Bs$     & $\mathcal{N} (-3.1, 0.3^2) $ & $\mathcal{N} (-3.1, 0.3^2) $  \\
$\ad$     & $\mathcal{U} (-1, 0) $ & $\mathcal{U} (-1, 0) $   \\
$\as$     & fixed & $\mathcal{U} (-1, 0) $\\
$\epsilon$ & fixed & $\mathcal{U} (0, 1) $ \\
\hline
\end{tabular}
\vspace*{2mm}
\label{tab:priors}
\vspace{0.1cm}
\end{table}

We also generate simulated lensing templates using only one of \sptpol, \bk, and \planck\ 
for the input $Q/U$ maps. 
We add the single-experiment lensing template to the BK14 simulation set
and perform maximum-likelihood searches. 
We find that the $\sigma(r)$ from  LT$_{\rm \sptpol}$, LT$_{\rm BICEP/Keck}$, and LT$_{\rm \planck}$
to be 0.0223, 0.0230, and 0.0236 respectively.\footnote{The 3-experiment Q/U combined LT gives $\sigma(r) = 0.0221$. 
We provide 3 significant figures for comparisons between the templates.}
This shows that the \sptpol\ $Q/U$ maps contribute most to recovering the lensing $B$-modes.
The fact that LT$_{\rm BICEP/Keck}$ contributes more than LT$_{\planck}$ 
shows that the signal-to-noise per mode at low $\ell$ is more important than
having a wider range in $\ell$ for the particular combination of the $\ell$ range
and noise levels between \bk\ and \planck.

\subsection{Parameter posteriors of BK14 with delensing}

\begin{figure*}
\begin{center}
\includegraphics[width=0.98\textwidth]{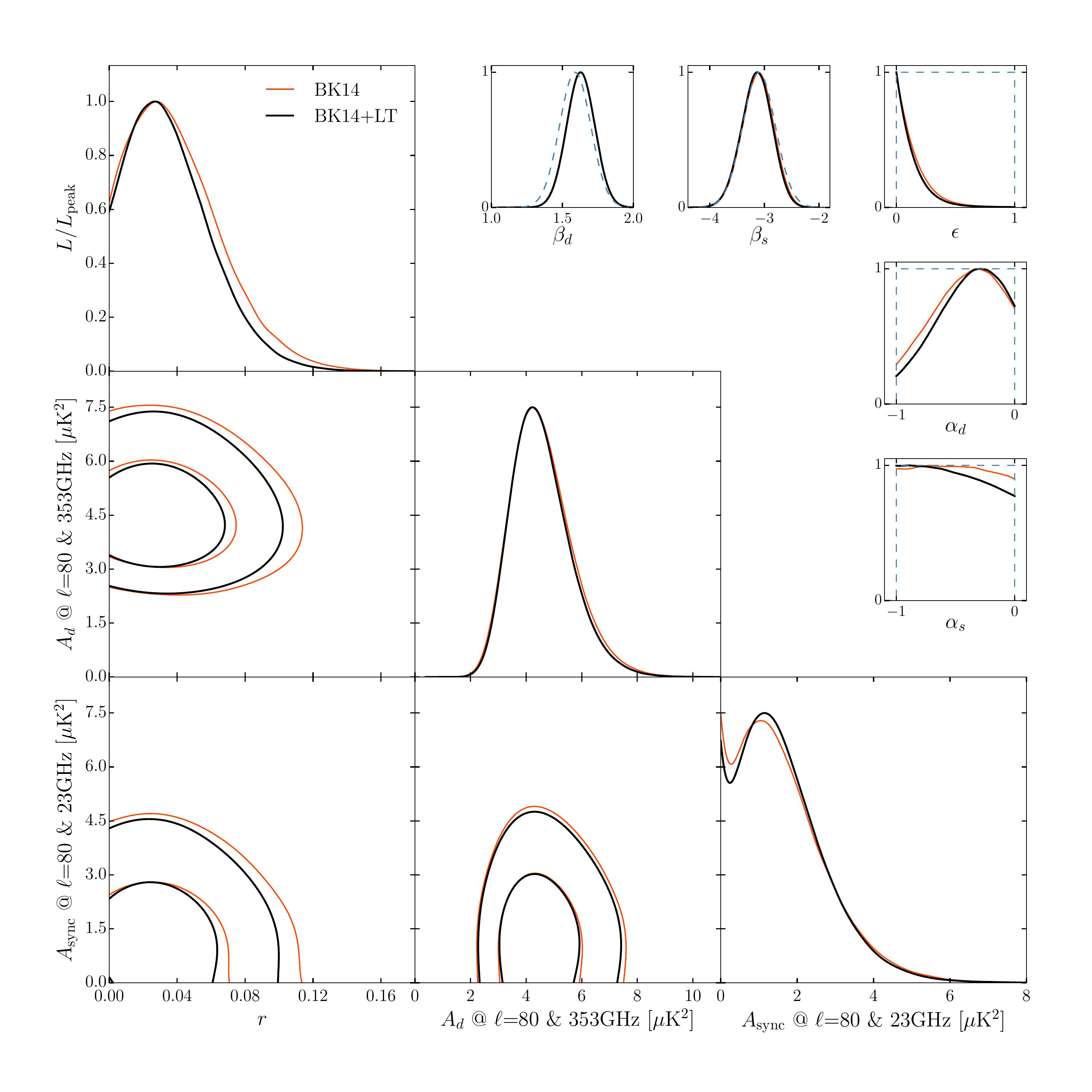}
\caption{
Posterior distributions of the baseline model parameters given the BK14+LT data set (black lines) 
compared with the BK14 data set (red lines, which are the same as the black lines in Fig. 4 of the BK14 paper). 
The lensing template is constructed using combined $Q/U$ maps from \sptpol,  \bk, and \planck\
(\refsec{combine_qu}) and a CIB map as the $\phi$ tracer (\refsec{cib_data}). 
The 95\% C.L. upper limit on the tensor-to-scalar ratio tightens from $r_{0.05} < \baselineoldRupperlim$ 
to $r_{0.05} < \baselineRupperlim$ with the addition of the lensing template.
The parameters $\Ad$ and $\As$ are the amplitudes of the dust and synchrotron 
B-mode spectra, where $\beta$ and $\alpha$ are the frequency and spatial spectral indices respectively.
The dust-synchrotron correlation parameter is denoted by $\epsilon$.
The up-turn of the 1D posterior distribution of $\As$ as it approaches zero comes
from the increased volume allowed by the $\epsilon$ parameter as $\epsilon$ becomes ambiguous when 
$\As=0$.
In the 1D panels for the $\alpha$, $\beta$, and $\epsilon$ parameters, the blue dashed lines denote
the priors for each parameter. 
}
\label{fig:triangle_baseline}
\end{center}
\end{figure*}

We now repeat the eight-parameter likelihood evaluation of the real data
as in the BK14 paper.
We again use \texttt{COSMOMC}~\citep{cosmomc} and the lensed-\lcdm+dust+synchrotron+$r$ model
with parameters and priors summarized in \reftab{priors}.
\reffig{triangle_baseline} shows the posterior distributions of the baseline analysis
compared with the BK14 result. 
The peak and 68\% credible regions of the marginalized $r$ distribution are shifted down from the BK14 values of \baselineoldRmarg\ 
to \baselineRmarg\ when the lensing template is included.
The 95\% C.L. upper limit on $r_{0.05}$ is reduced from \baselineoldRupperlim\  to \baselineRupperlim.
Some of the other constraints are $\Ad =  \baselineAdmarg$\,\uksq\ and $\As <  \baselineAsupperlim$\,\uksq\ (95\% C.L.).\footnote{
As noted, the model space is identical to BK14 to enable apples-to-apples comparison. 
However, we have since then made one model change in BK15~\cite{bk10} and widened the prior
range of the dust-synchrotron correlation parameter $\epsilon$ from $0 < \epsilon < 1$ to 
$-1 < \epsilon < 1$ (see Appendix E1 in BK15 for details). With this prior, the BK14 $r$
peak and 68\% credible regions reduce from \oldepsRmarg\ to \epsRmarg\ when a lensing template is
included.}
The maximum-likelihood model (including priors) in the 8D parameter space is:
$r_{0.05} = 0.025$,
$\Ad = 4.0$\,\uksq, 
$\As = 1.4$\,\uksq, 
$\Bd = 1.6$, $\Bs = -3.1$, 
$\ad = -0.17$, $\as = -0.95$,
and $\epsilon= 0.00$.
Against this model,
we compute $\chi^2 = (d-m)^{\dagger} {\rm Cov^{-1}} (d-m) = 768$ for
the $ 9 \times 78 = 702$ data bandpowers.
We compare this number against the distribution in simulations finding a PTE of 0.15.
We conclude that the model is a sufficient description of the data at present.

We perform a couple of variations to the baseline analysis to explore
degeneracies amongst model parameters that are important to lensing,
and changes in $r$ with different input data sets.
In the baseline analysis, the lensing $BB$ spectrum is taken
as the \lcdm\ expectation in both normalization and shape.
As an alternative we re-scale this spectrum by the parameter
$A_{\rm L}$ and sample the posterior 
distribution in the \lcdm+$A_{\rm L}$ model space. 
Secondly, as is done in~\refsec{sim_sr}, we form input lensing templates 
using $Q/U$ maps from one of the three experiments instead of combining them.
We discuss the results of each variation in the following paragraphs.

\begin{figure}
\begin{center}
\includegraphics[width=0.48\textwidth]{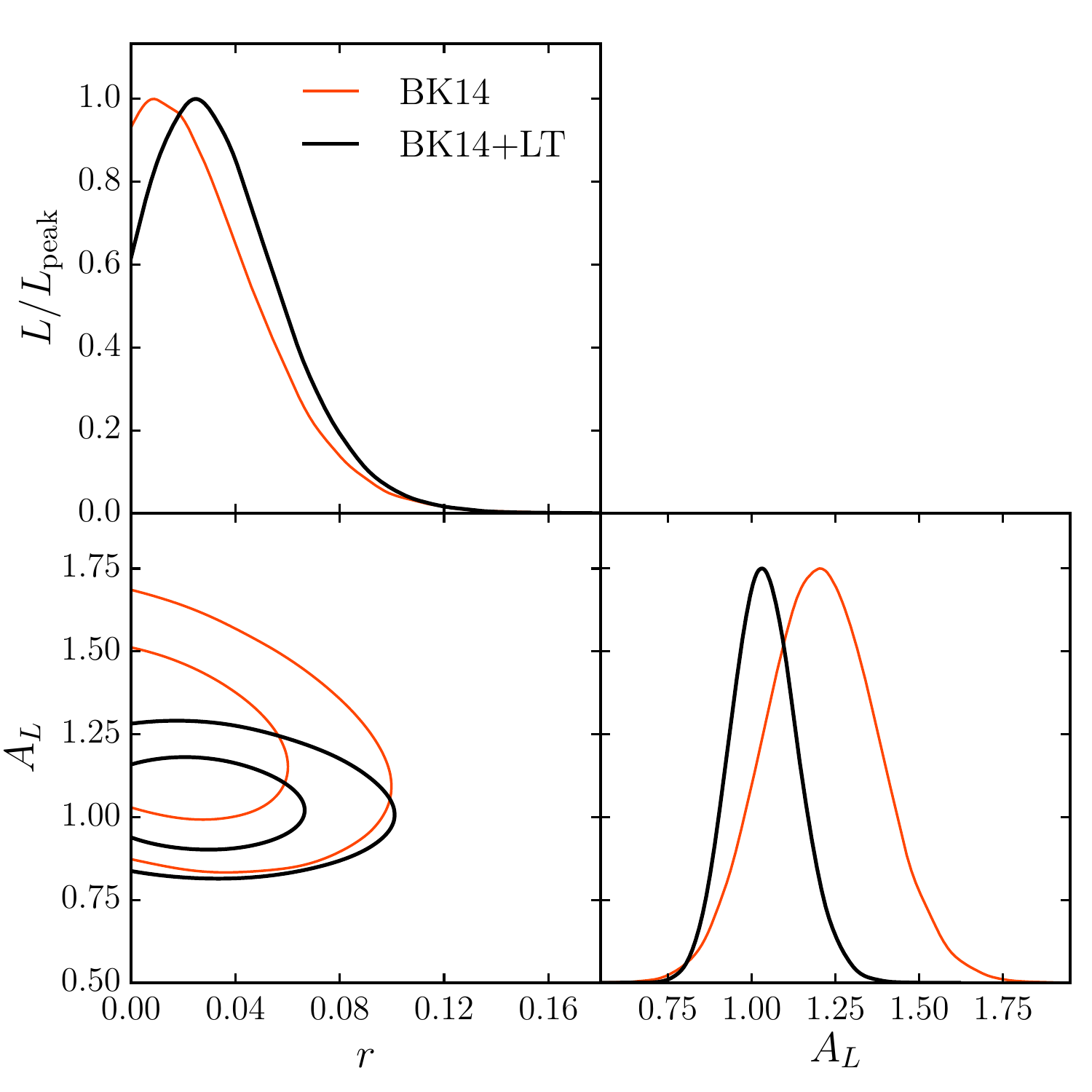}
\caption{Posterior distributions on $r$ and $A_{\rm L}$, a parameter used to scale the lensing $BB$ power, 
from an alternative analysis in which the amplitude of lensing is a free parameter.
With the addition of the lensing template, the probability of shuffling lensing power 
to other parameters is reduced,
thus the degeneracy between $r$ and $A_{\rm L}$ is reduced.}
\label{fig:alens}
\end{center}
\end{figure}

When we allow $A_{\rm L}$ to float, we note a $A_{\rm L}-r$ degeneracy in the BK14 data set,
as shown in~\reffig{alens}, and as was previously noted in an earlier 
\bk\ analysis~\cite{bkp}.
When the lensing template is added to the BK14 data set, the degeneracy
between $r$ and $A_{\rm L}$ is reduced.
In this model space, the peak and 68\% credible regions of the marginalized $r$ distribution
 with and without the lensing template are \alensRmarg\ and \alensoldRmarg, 
and the upper limits on $r$ are $r_{0.05} < \alensRupperlim$ and $r_{0.05} < \alensoldRupperlim$ respectively.
The peak and 68\% credible regions of $A_{\rm L}$ with and without the lensing template 
are \alensALmarg\ and \alensoldALmarg\ respectively.\footnote{We note that 
we have kept fixed a component of the noise bias in the LT auto-spectrum ($s_\phi \ast n_{QU}$ in ~\refeqn{lt_noise_spec})
which varies with $A_{\rm L}$. It contributes $<10\%$ of the total noise bias and is
only present in the LT$\times$LT part of the data vector. 
Varying this noise component with $A_{\rm L}$ would slightly tighten the constraint on $A_{\rm L}$, but the
qualitative conclusion would be changed. }
The shift in the peak $A_{\rm L}$ is consistent with expectations from simulations, where 25\% of the
simulation realizations have $A_{\rm L}$ shifts with absolute magnitude larger than that seen in data. 
We see that with the addition of the lensing template, we are able to better constrain the lensing 
power in the measured auto- and cross-spectra across the different frequencies
and thereby reduce the probability of mis-assigning power to lensing.

\begin{figure}
\begin{center}
\includegraphics[width=0.43\textwidth]{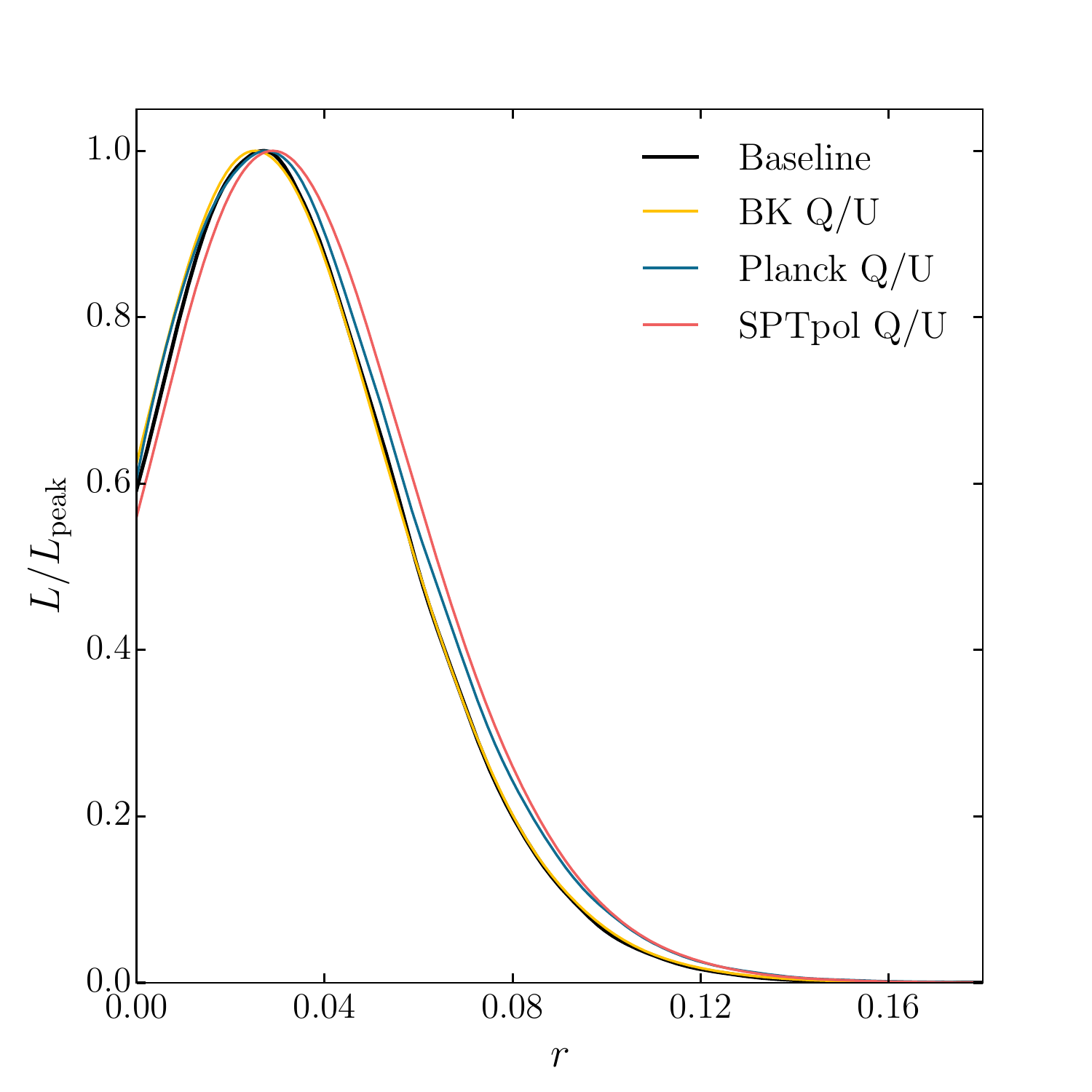}
\caption{The $r$ posterior curves from the baseline analysis, along with 
$r$ curves from analyses using lensing templates constructed from $Q/U$ maps 
from only one of the three experiments:  \bk, \planck, and \sptpol.
The shifts in the curves are consistent with expectations from simulations.}
\label{fig:r_indivqu}
\end{center}
\end{figure}

We show in~\reffig{r_indivqu} the $r$ posterior distributions from analyses 
in the lensed-\lcdm\ model space using lensing templates constructed from $Q/U$ maps coming 
from only one of the three experiments, \sptpol, \bk\ and \planck. 
We see that the peaks of the $r$ posteriors from the \bk-only and the \planck-only cases
are close to the baseline case, while the width of the $r$ posterior from the \planck-only case
is a bit larger than the baseline case.
The larger $r$ posterior uncertainty is expected given the larger $\sigma(r)$ from the \planck-only
simulation set in~\refsec{sim_sr}.
The peak of the $r$ posterior for 
the \sptpol-only case is shifted up slightly compared with the baseline case. 
This might seem slightly surprising 
given that the \sptpol\ $Q/U$ maps contribute most of the weight in the combined $Q/U$ maps over
a broad range of angular scales. 
To quantify the probability of the observed shift between the baseline case and the \sptpol-only case,
we extract the best-fit $r$ values from the baseline simulation set
and the \sptpol-only lensing template simulation set. 
Restricting to the subset with positive best-fit $r$ in the baseline setup, we count the fraction of
realizations that have larger best-fit $r$ differences between the \sptpol-only and the baseline set
than is seen in the data.
We find 20\% of the simulations fit this criterion and 
thus we conclude that what is observed in the data is typical of the expected fluctuations.

\section{Conclusion}\label{sec:conclusion}
In this work, we build on the \bk\ analysis framework
and demonstrate, for the first time, improvements to constraints on 
the tensor-to-scalar ratio $r$ with delensing.
With the addition of a lensing template, we reduce the uncertainty of the $r$ estimate
by constraining the lensing $B$-mode contribution to the observed $B$-modes. 
We construct the lensing template using an undeflect-and-difference approach,
in which we undeflect the observed $Q/U$ maps by a $\phi$ tracer,
and then difference the undeflected maps from the input maps.
The $Q/U$ maps we use are a 150\,GHz combination of \sptpol\ observations from 2013--2015,
\bk\ observations up to 2014, and the \planck\ satellite full-mission
observations.
The $\phi$ tracer we use is a CIB map constructed using the GNILC algorithm 
from \planck\ data.
The resulting lensing template is added as a pseudo-frequency band to the BK14 dataset, in which \bk\,
 \wmap\, and \planck\ maps are used to constrain Galactic foregrounds and $r$.

We present two key results from this analysis. 
First, we estimate $\sigma(r)$ using our lensed-\lcdm+dust+noise simulation set.
We find maximum likelihood values of the baseline model parameters for each simulation realization
and take the mean and standard deviation over the 499 realizations.
We find that, with the addition of the lensing template, 
$\sigma(r)$ improves from \simbksr\ in BK14 to \simbkltsr, a $\sim 10\%$ improvement.
The second main result is the posterior peak value, 68\% credible region, and upper limit on $r$
when we add the lensing template to the BK14 data set. 
With delensing,
the peak and 68\% credible regions shift from $r = \baselineoldRmarg$ to $r = \baselineRmarg$, and 
the 95\% C.L. upper limit on $r$ is reduced from $\baselineoldRupperlim$ to $\baselineRupperlim$. 

We estimate the impact on $r$ from potential biases in the inputs used to construct
the simulated lensing templates.
We find the biases to $r$ from misestimating the cross-spectrum of the CIB and $\phi$ to be small,
and the biases to $r$ from biases in polarization efficiency of the CMB $Q/U$ maps to be negligible.
We find negligible difference in $\sigma(r)$ due to modeling the non-Gaussian CIB field as Gaussian
for this data set, 
and that the uncertainties in the CIB auto-spectrum and the CIB$\times\phi$ cross-spectrum
contribute sub-dominantly to $\sigma(r)$. 
We perform checks against potential unmodeled systematic contaminations to the
lensing template.
This includes Galactic foregrounds leaking into the lensing template through either the
input $Q/U$ maps or the input CIB map.
We show that the data lensing template is sufficiently well-described by the simulations.
Therefore we conclude that the results are robust against these sources of systematics
given the current noise levels.

At the BK14 level of map noise and Galactic foreground variance,
simulations indicate that perfect delensing would reduce $\sigma(r)$ from \simbksr\ to \simbkperfectltsr.
This implies that the variance from lensing $B$-modes is not the dominant source of
uncertainty ($< 30\%$) when constraining $r$ in this data set. 
However, with current and upcoming ground-based CMB telescopes, e.g. \bicep\ Array~\cite{biceparray20},
SPT-3G~\cite{bender18}, AdvACT, Simons Array, Simons Observatory~\cite{SO18}, and CMB-S4~\cite{cmbs4-sb1}, the millimeter-wave
sky will be mapped with ever higher signal-to-noise. 
Lensing $B$-modes will become a dominant source of uncertainty,
and delensing will be crucial to break the floor of $\sigma(r)$ set by the lensing variance.
For example, while in the most recent \bk\ $r$ analysis BK15~\cite{bk10} lensing variance continues to be
subdominant, in the upcoming result BK18 lensing variance contributes roughly half of the $r$
uncertainty budget. 
Projecting further, without delensing, the \bicep\ Array experiment $\sigma(r)$ would plateau at $\sim$0.006.
However, this $\sigma(r)$ could be reduced by a factor of about 2.5
with delensing using a $\phi$ field reconstructed using CMB maps from the SPT-3G experiment.
This is a much more significant reduction in the uncertainty on $r$ 
than is achieved in this work.

To reach the target $\sigma(r)$ of $5\times 10^{-4}$ for the next-generation ground-based
CMB experiment CMB-S4, more than 90\% of the lensing sample variance
needs to be removed~\cite{s4pgw}. 
Delensing to such low residual levels requires 
high values of $\rho_{L}$, the correlation between the $\phi$ tracer and the underlying $\phi$ field.
In addition to 
using $\phi$ maps reconstructed from
low-noise, high-resolution CMB observations~\cite[e.g.][]{wu2019, pblens2019, actlens}, 
higher $\rho_{L}$ tracers could be obtained by 
combining different tracers~\cite[e.g.][]{manzotti18, yu17} and
using optimal methods~\cite[e.g.][]{seljakhirata04, lensit, bayeslens, caldeira}.
We will be exploring various approaches to delensing~\cite[e.g.][]{bayeslens} in future joint analyses of
\bk\ and SPT-3G data,  
confronting delensing algorithms with real-world non-idealities
and developing techniques to mitigate systematics, readying our analysis for
the future of low-noise data and the possibility of detecting PGWs.

\begin{appendix}

\section{Lensing template construction methods}
\label{app:templates}

In this paper, we have used a map-space ``undeflect-and-difference" method
to construct the lensing template.
Previous works have inferred the lensing $B$-modes $\bhat(\bl)$
by lensing the observed $E$-modes to first order in $\phi$ given a $\phi$ tracer. 
Specifically,  
\begin{equation}
\bhat(\bl) = \int \frac{d^2 \bl'}{(2 \pi)^2} W(\bl,\bl') \bar{\phi}(\bl - \bl') \bar{E}(\ublu)  \,,
\label{eqn:fspacelens}
\end{equation}
where
$W(\bl,\bl') = \bl' \cdot (\bl-\bl') \sin(2\varphi_{\bl,\bl'})$, and
$\bar{E}$ and $ \bar{\phi}$ are the Wiener filtered $E$-modes and $\phi$ tracer
respectively~\cite[e.g.][]{manzotti17}.
An advantage to this formulation is that by acting on the
$E$-modes of the observed sky only, noise in the lensing template is reduced versus the
undeflect-and-difference method. 
This extra noise enters by undeflecting $Q/U$ maps which also contain $B$-modes.
While the signal contribution from the $B$-modes is small, the level of noise fluctuations 
is similar to those in the $E$-modes, which contribute noise to the 
undeflect-and-difference templates. 
However, this is not a fundamental limitation to the map-space approach 
as implemented in this paper. 
One could Fourier transform
the $Q/U$ maps to $E/B$-modes, null the $B$-modes, and then transform
back to $Q/U$ maps before performing the undeflect-and-difference operation, 
which would remove this specific noise. 
In fact, we experimented with adding these steps and found that for the
present case, the reduction in lensing template noise is fractionally
very small for $\ell < 500$.

For future analyses, we will revisit the algorithm used to produce the lensing template
to further improve its signal-to-noise.
Besides removing the extra noise contribution, 
other possible improvements include filling in the region outside the \sptpol\ coverage  (as seen in \reffig{exptcomb})
using the information available from \bk\ and \planck.

\end{appendix}

\begin{acknowledgements}
The authors thank Dominic Beck and Chang Feng for useful comments on an early version of the draft. 
The \biceptwo/\keckarray\ projects have been made possible through
a series of grants from the National Science Foundation
including 0742818, 0742592, 1044978, 1110087, 1145172, 1145143, 1145248,
1639040, 1638957, 1638978, \& 1638970, and by the Keck Foundation.
The development of antenna-coupled detector technology was supported
by the JPL Research and Technology Development Fund, and by NASA Grants
06-ARPA206-0040, 10-SAT10-0017, 12-SAT12-0031, 14-SAT14-0009
\& 16-SAT-16-0002.
The development and testing of focal planes were supported
by the Gordon and Betty Moore Foundation at Caltech.
Readout electronics were supported by a Canada Foundation
for Innovation grant to UBC.
Support for quasi-optical filtering was provided by UK STFC grant ST/N000706/1.
Some of the computations in this paper were run on the Odyssey cluster
supported by the FAS Science Division Research Computing Group at
Harvard University.
The analysis effort at Stanford and SLAC is partially supported by
the U.S. DOE Office of Science.
We thank the staff of the U.S. Antarctic Program and in particular
the South Pole Station without whose help this research would not
have been possible.
Most special thanks go to our heroic winter-overs Robert Schwarz
and Steffen Richter.
We thank all those who have contributed past efforts to the \bicep--\keckarray\
series of experiments, including the \bicepone\ team.

SPT is supported by the National Science Foundation through grants PLR-1248097 and OPP-1852617.
Partial support is also provided by the NSF Physics Frontier Center grant PHY-1125897 to the Kavli Institute of Cosmological Physics at the University of Chicago, the Kavli Foundation and the Gordon and Betty Moore Foundation grant GBMF 947. This research used resources of the National Energy Research Scientific Computing Center (NERSC), a DOE Office of Science User Facility supported by the Office of Science of the U.S. Department of Energy under Contract No. DE-AC02-05CH11231.  
The Melbourne group acknowledges support from the University of Melbourne and an Australian Research Council's Future Fellowship (FT150100074). 
Work at Argonne National Lab is supported by UChicago Argonne LLC, Operator  of  Argonne  National  Laboratory  (Argonne). Argonne, a U.S. Department of Energy Office of Science Laboratory,  is  operated  under  contract  no.   DE-AC02-06CH11357.  We also acknowledge support from the Argonne  Center  for  Nanoscale  Materials.  
Work at McGill is supported by the Natural Science and Engineering Research Council of Canada, the Canadian Institute for Advanced Research, and M.D. acknowledges a Killam research fellowship 
W.L.K.W is supported in part by the Kavli Institute for Cosmological Physics at the University of Chicago through grant NSF PHY-1125897, an endowment from the Kavli Foundation and its founder Fred Kavli, and by the Department of Energy, Laboratory Directed Research and Development program and as part of the Panofsky Fellowship program at SLAC National Accelerator Laboratory, under contract DE-AC02-76SF00515.
B.B. is supported by the Fermi Research Alliance LLC under contract no. De-AC02- 07CH11359 with the U.S. Department of Energy.

We acknowledge the use of many python packages: {\sc IPython} \citep{ipython}, {\sc matplotlib} \citep{matplotlib}, {\sc scipy} \citep{scipy}, and {\sc healpy} \citep{healpy, healpix}.
We also thank the \planck\ and \wmap\ teams for the use of their data.
Some of the sky simulations used in this paper were developed by the WebSky Extragalactic CMB Mocks team, with the continuous support of the Canadian Institute for Theoretical Astrophysics (CITA), the Canadian Institute for Advanced Research (CIFAR), and the Natural Sciences and Engineering Research Council of Canada (NSERC), and were generated on the Niagara supercomputer at the SciNet HPC Consortium. SciNet is funded by: the Canada Foundation for Innovation under the auspices of Compute Canada; the Government of Ontario; Ontario Research Fund - Research Excellence; and the University of Toronto.
\end{acknowledgements}

\bibliographystyle{natbib/utphys_etal}
\bibliography{ms}

\providecommand{\href}[2]{#2}\begingroup\raggedright\begin{thebibliography}{10}

\bibitem{kamionkowski2016}
M.~{Kamionkowski} and E.~D. {Kovetz}, ``{The Quest for B Modes from
  Inflationary Gravitational Waves},''
  \href{http://dx.doi.org/10.1146/annurev-astro-081915-023433}{{\em \araa}
  {\bfseries 54} (Sept., 2016) 227--269},
  \href{http://arxiv.org/abs/1510.06042}{{\ttfamily arXiv:1510.06042
  [astro-ph.CO]}}.

\bibitem{planck2018inflation}
{Planck Collaboration}, Y.~{Akrami}, F.~{Arroja}, {\em et~al.}, ``{Planck 2018
  results. X. Constraints on inflation},'' {\em arXiv e-prints} (July, 2018)
  arXiv:1807.06211, \href{http://arxiv.org/abs/1807.06211}{{\ttfamily
  arXiv:1807.06211 [astro-ph.CO]}}.

\bibitem{seljak97}
U.~{Seljak} and M.~{Zaldarriaga}, ``{Signature of Gravity Waves in the
  Polarization of the Microwave Background},''
  \href{http://dx.doi.org/10.1103/PhysRevLett.78.2054}{{\em \prl} {\bfseries
  78} no.~11, (Mar., 1997) 2054--2057},
  \href{http://arxiv.org/abs/astro-ph/9609169}{{\ttfamily
  arXiv:astro-ph/9609169 [astro-ph]}}.

\bibitem{kamion97}
M.~{Kamionkowski}, A.~{Kosowsky}, and A.~{Stebbins}, ``{A Probe of Primordial
  Gravity Waves and Vorticity},''
  \href{http://dx.doi.org/10.1103/PhysRevLett.78.2058}{{\em \prl} {\bfseries
  78} no.~11, (Mar., 1997) 2058--2061},
  \href{http://arxiv.org/abs/astro-ph/9609132}{{\ttfamily
  arXiv:astro-ph/9609132 [astro-ph]}}.

\bibitem{planckintXXX}
{Planck Collaboration}, R.~{Adam}, P.~A.~R. {Ade}, {\em et~al.}, ``{Planck
  intermediate results. XXX. The angular power spectrum of polarized dust
  emission at intermediate and high Galactic latitudes},''
  \href{http://dx.doi.org/10.1051/0004-6361/201425034}{{\em \aap} {\bfseries
  586} (Feb., 2016) A133}, \href{http://arxiv.org/abs/1409.5738}{{\ttfamily
  arXiv:1409.5738 [astro-ph.CO]}}.

\bibitem{spass18}
N.~{Krachmalnicoff}, E.~{Carretti}, C.~{Baccigalupi}, {\em et~al.}, ``{S-PASS
  view of polarized Galactic synchrotron at 2.3 GHz as a contaminant to CMB
  observations},'' \href{http://dx.doi.org/10.1051/0004-6361/201832768}{{\em
  \aap} {\bfseries 618} (Oct., 2018) A166},
  \href{http://arxiv.org/abs/1802.01145}{{\ttfamily arXiv:1802.01145
  [astro-ph.GA]}}.

\bibitem{lewis2006}
A.~{Lewis} and A.~{Challinor}, ``{Weak gravitational lensing of the CMB},''
  \href{http://dx.doi.org/10.1016/j.physrep.2006.03.002}{{\em \physrep}
  {\bfseries 429} (June, 2006) 1--65},
  \href{http://arxiv.org/abs/astro-ph/0601594}{{\ttfamily astro-ph/0601594}}.

\bibitem{Hanson:2013hsb}
D.~{Hanson}, S.~{Hoover}, A.~{Crites}, {\em et~al.}, ``{Detection of B-Mode
  Polarization in the Cosmic Microwave Background with Data from the South Pole
  Telescope},'' \href{http://dx.doi.org/10.1103/PhysRevLett.111.141301}{{\em
  \prl} {\bfseries 111} no.~14, (Oct., 2013) 141301},
  \href{http://arxiv.org/abs/1307.5830}{{\ttfamily arXiv:1307.5830
  [astro-ph.CO]}}.

\bibitem{polarbearbb14}
{Polarbear Collaboration}, {P.~A.~R.~Ade}, Y.~{Akiba}, {\em et~al.}, ``{A
  Measurement of the Cosmic Microwave Background B-mode Polarization Power
  Spectrum at Sub-degree Scales with POLARBEAR},''
  \href{http://dx.doi.org/10.1088/0004-637X/794/2/171}{{\em \apj} {\bfseries
  794} (Oct., 2014) 171}, \href{http://arxiv.org/abs/1403.2369}{{\ttfamily
  arXiv:1403.2369}}.

\bibitem{keisler15}
R.~{Keisler}, S.~{Hoover}, N.~{Harrington}, {\em et~al.}, ``{Measurements of
  Sub-degree B-mode Polarization in the Cosmic Microwave Background from 100
  Square Degrees of SPTpol Data},''
  \href{http://dx.doi.org/10.1088/0004-637X/807/2/151}{{\em \apj} {\bfseries
  807} (July, 2015) 151}.

\bibitem{louis16}
T.~{Louis}, E.~{Grace}, M.~{Hasselfield}, {\em et~al.}, ``{The Atacama
  Cosmology Telescope: two-season ACTPol spectra and parameters},''
  \href{http://dx.doi.org/10.1088/1475-7516/2017/06/031}{{\em \jcap} {\bfseries
  6} (June, 2017) 031}, \href{http://arxiv.org/abs/1610.02360}{{\ttfamily
  arXiv:1610.02360}}.

\bibitem{bk6}
{BICEP2 Collaboration}, {Keck Array Collaboration}, P.~A.~R. {Ade}, {\em
  et~al.}, ``{Improved Constraints on Cosmology and Foregrounds from BICEP2 and
  Keck Array Cosmic Microwave Background Data with Inclusion of 95 GHz Band},''
  \href{http://dx.doi.org/10.1103/PhysRevLett.116.031302}{{\em Physical Review
  Letters} {\bfseries 116} no.~3, (Jan., 2016) 031302},
  \href{http://arxiv.org/abs/1510.09217}{{\ttfamily arXiv:1510.09217}}.

\bibitem{polarbearbb17}
{POLARBEAR Collaboration}, P.~A.~R. {Ade}, M.~{Aguilar}, {\em et~al.}, ``{A
  Measurement of the Cosmic Microwave Background B-mode Polarization Power
  Spectrum at Subdegree Scales from Two Years of polarbear Data},''
  \href{http://dx.doi.org/10.3847/1538-4357/aa8e9f}{{\em \apj} {\bfseries 848}
  (Oct., 2017) 121}, \href{http://arxiv.org/abs/1705.02907}{{\ttfamily
  arXiv:1705.02907}}.

\bibitem{bk10}
{BICEP2 Collaboration}, {Keck Array Collaboration}, P.~A.~R. {Ade}, {\em
  et~al.}, ``{Constraints on Primordial Gravitational Waves Using Planck, WMAP,
  and New BICEP2/Keck Observations through the 2015 Season},''
  \href{http://dx.doi.org/10.1103/PhysRevLett.121.221301}{{\em \prl} {\bfseries
  121} no.~22, (Nov, 2018) 221301},
  \href{http://arxiv.org/abs/1810.05216}{{\ttfamily arXiv:1810.05216
  [astro-ph.CO]}}.

\bibitem{bkp}
{BICEP2/Keck Collaboration}, {Planck Collaboration}, P.~A.~R. {Ade}, {\em
  et~al.}, ``{Joint Analysis of BICEP2/Keck Array and Planck Data},''
  \href{http://dx.doi.org/10.1103/PhysRevLett.114.101301}{{\em \prl} {\bfseries
  114} no.~10, (Mar., 2015) 101301},
  \href{http://arxiv.org/abs/1502.00612}{{\ttfamily arXiv:1502.00612
  [astro-ph.CO]}}.

\bibitem{manzotti17}
A.~{Manzotti}, K.~T. {Story}, W.~L.~K. {Wu}, {\em et~al.}, ``{CMB Polarization
  B-mode Delensing with SPTpol and Herschel},''
  \href{http://dx.doi.org/10.3847/1538-4357/aa82bb}{{\em \apj} {\bfseries 846}
  (Sept., 2017) 45}, \href{http://arxiv.org/abs/1701.04396}{{\ttfamily
  arXiv:1701.04396}}.

\bibitem{carron17}
J.~{Carron}, A.~{Lewis}, and A.~{Challinor}, ``{Internal delensing of Planck
  CMB temperature and polarization},''
  \href{http://dx.doi.org/10.1088/1475-7516/2017/05/035}{{\em \jcap} {\bfseries
  5} (May, 2017) 035}, \href{http://arxiv.org/abs/1701.01712}{{\ttfamily
  arXiv:1701.01712}}.

\bibitem{plancklens18}
{Planck Collaboration}, N.~{Aghanim}, Y.~{Akrami}, {\em et~al.}, ``{Planck 2018
  results. VIII. Gravitational lensing},'' {\em arXiv e-prints} (Jul, 2018)
  arXiv:1807.06210, \href{http://arxiv.org/abs/1807.06210}{{\ttfamily
  arXiv:1807.06210 [astro-ph.CO]}}.

\bibitem{pbdelens19}
S.~{Adachi}, M.~A.~O. {Aguilar Fa{\'u}ndez}, Y.~{Akiba}, {\em et~al.},
  ``{Internal delensing of cosmic microwave background polarization B-modes
  with the POLARBEAR experiment},'' {\em arXiv e-prints} (Sep, 2019)
  arXiv:1909.13832, \href{http://arxiv.org/abs/1909.13832}{{\ttfamily
  arXiv:1909.13832 [astro-ph.CO]}}.

\bibitem{actdelens20}
D.~{Han}, N.~{Sehgal}, A.~{MacInnis}, {\em et~al.}, ``{The Atacama Cosmology
  Telescope: Delensed Power Spectra and Parameters},'' {\em arXiv e-prints}
  (July, 2020) arXiv:2007.14405,
  \href{http://arxiv.org/abs/2007.14405}{{\ttfamily arXiv:2007.14405
  [astro-ph.CO]}}.

\bibitem{huokamoto02}
W.~{Hu} and T.~{Okamoto}, ``{Mass Reconstruction with Cosmic Microwave
  Background Polarization},'' \href{http://dx.doi.org/10.1086/341110}{{\em
  \apj} {\bfseries 574} no.~2, (Aug., 2002) 566--574},
  \href{http://arxiv.org/abs/astro-ph/0111606}{{\ttfamily
  arXiv:astro-ph/0111606 [astro-ph]}}.

\bibitem{simard15}
G.~{Simard}, D.~{Hanson}, and G.~{Holder}, ``{Prospects for Delensing the
  Cosmic Microwave Background for Studying Inflation},''
  \href{http://dx.doi.org/10.1088/0004-637X/807/2/166}{{\em \apj} {\bfseries
  807} no.~2, (Jul, 2015) 166},
  \href{http://arxiv.org/abs/1410.0691}{{\ttfamily arXiv:1410.0691
  [astro-ph.CO]}}.

\bibitem{sherwin15}
B.~D. {Sherwin} and M.~{Schmittfull}, ``{Delensing the CMB with the cosmic
  infrared background},''
  \href{http://dx.doi.org/10.1103/PhysRevD.92.043005}{{\em \prd} {\bfseries 92}
  no.~4, (Aug., 2015) 043005},
  \href{http://arxiv.org/abs/1502.05356}{{\ttfamily arXiv:1502.05356
  [astro-ph.CO]}}.

\bibitem{wu2019}
W.~L.~K. {Wu}, L.~M. {Mocanu}, P.~A.~R. {Ade}, {\em et~al.}, ``{A Measurement
  of the Cosmic Microwave Background Lensing Potential and Power Spectrum from
  500 deg$^{2}$ of SPTpol Temperature and Polarization Data},''
  \href{http://dx.doi.org/10.3847/1538-4357/ab4186}{{\em \apj} {\bfseries 884}
  no.~1, (Oct., 2019) 70}, \href{http://arxiv.org/abs/1905.05777}{{\ttfamily
  arXiv:1905.05777 [astro-ph.CO]}}.

\bibitem{planck2015cibdust}
{Planck Collaboration}, N.~{Aghanim}, M.~{Ashdown}, {\em et~al.}, ``{Planck
  intermediate results. XLVIII. Disentangling Galactic dust emission and cosmic
  infrared background anisotropies},''
  \href{http://dx.doi.org/10.1051/0004-6361/201629022}{{\em \aap} {\bfseries
  596} (Dec., 2016) A109}, \href{http://arxiv.org/abs/1605.09387}{{\ttfamily
  arXiv:1605.09387 [astro-ph.CO]}}.

\bibitem{plancklens15}
{Planck Collaboration}, P.~A.~R. {Ade}, N.~{Aghanim}, {\em et~al.}, ``{Planck
  2015 results. XV. Gravitational lensing},''
  \href{http://dx.doi.org/10.1051/0004-6361/201525941}{{\em \aap} {\bfseries
  594} (Sept., 2016) A15}, \href{http://arxiv.org/abs/1502.01591}{{\ttfamily
  arXiv:1502.01591}}.

\bibitem{hu2000}
W.~{Hu}, ``{Weak lensing of the CMB: A harmonic approach},''
  \href{http://dx.doi.org/10.1103/PhysRevD.62.043007}{{\em \prd} {\bfseries 62}
  (Aug., 2000) }.

\bibitem{anderes15}
E.~{Anderes}, B.~D. {Wandelt}, and G.~{Lavaux}, ``{Bayesian Inference of CMB
  Gravitational Lensing},''
  \href{http://dx.doi.org/10.1088/0004-637X/808/2/152}{{\em \apj} {\bfseries
  808} no.~2, (Aug., 2015) 152},
  \href{http://arxiv.org/abs/1412.4079}{{\ttfamily arXiv:1412.4079
  [astro-ph.CO]}}.

\bibitem{green16}
D.~{Green}, J.~{Meyers}, and A.~{van Engelen}, ``{CMB delensing beyond the B
  modes},'' \href{http://dx.doi.org/10.1088/1475-7516/2017/12/005}{{\em \jcap}
  {\bfseries 12} (Dec., 2017) 005},
  \href{http://arxiv.org/abs/1609.08143}{{\ttfamily arXiv:1609.08143}}.

\bibitem{bk1}
{BICEP2 Collaboration}, P.~A.~R. {Ade}, R.~W. {Aikin}, {\em et~al.},
  ``{Detection of B-Mode Polarization at Degree Angular Scales by BICEP2},''
  \href{http://dx.doi.org/10.1103/PhysRevLett.112.241101}{{\em Physical Review
  Letters} {\bfseries 112} no.~24, (June, 2014) 241101},
  \href{http://arxiv.org/abs/1403.3985}{{\ttfamily arXiv:1403.3985}}.

\bibitem{hl2008}
S.~{Hamimeche} and A.~{Lewis}, ``{Likelihood analysis of CMB temperature and
  polarization power spectra},''
  \href{http://dx.doi.org/10.1103/PhysRevD.77.103013}{{\em \prd} {\bfseries 77}
  no.~10, (May, 2008) 103013}, \href{http://arxiv.org/abs/0801.0554}{{\ttfamily
  arXiv:0801.0554}}.

\bibitem{austermann12}
J.~E. {Austermann}, K.~A. {Aird}, J.~A. {Beall}, {\em et~al.},
  \href{http://dx.doi.org/10.1117/12.927286}{``{SPTpol: an instrument for CMB
  polarization measurements with the South Pole Telescope},''} in {\em
  Millimeter, Submillimeter, and Far-Infrared Detectors and Instrumentation for
  Astronomy VI}, vol.~8452 of {\em Proc.~SPIE}, p.~84521E.
\newblock Sept., 2012.
\newblock \href{http://arxiv.org/abs/1210.4970}{{\ttfamily arXiv:1210.4970
  [astro-ph.IM]}}.

\bibitem{carlstrom11}
J.~E. {Carlstrom}, P.~A.~R. {Ade}, K.~A. {Aird}, {\em et~al.}, ``{The 10 Meter
  South Pole Telescope},'' \href{http://dx.doi.org/10.1086/659879}{{\em PASP}
  {\bfseries 123} (May, 2011) 568},
  \href{http://arxiv.org/abs/0907.4445}{{\ttfamily arXiv:0907.4445
  [astro-ph.IM]}}.

\bibitem{henning18}
J.~W. {Henning}, J.~T. {Sayre}, C.~L. {Reichardt}, {\em et~al.},
  ``{Measurements of the Temperature and E-mode Polarization of the CMB from
  500 Square Degrees of SPTpol Data},''
  \href{http://dx.doi.org/10.3847/1538-4357/aa9ff4}{{\em \apj} {\bfseries 852}
  (Jan., 2018) 97}, \href{http://arxiv.org/abs/1707.09353}{{\ttfamily
  arXiv:1707.09353}}.

\bibitem{smith2012}
K.~M. {Smith}, D.~{Hanson}, M.~{LoVerde}, C.~M. {Hirata}, and O.~{Zahn},
  ``{Delensing CMB polarization with external datasets},''
  \href{http://dx.doi.org/10.1088/1475-7516/2012/06/014}{{\em \jcap} {\bfseries
  2012} no.~6, (June, 2012) 014},
  \href{http://arxiv.org/abs/1010.0048}{{\ttfamily arXiv:1010.0048
  [astro-ph.CO]}}.

\bibitem{lenz19}
D.~{Lenz}, O.~{Dor{\'e}}, and G.~{Lagache}, ``{Large-scale Maps of the Cosmic
  Infrared Background from Planck},''
  \href{http://dx.doi.org/10.3847/1538-4357/ab3c2b}{{\em \apj} {\bfseries 883}
  no.~1, (Sep, 2019) 75}, \href{http://arxiv.org/abs/1905.00426}{{\ttfamily
  arXiv:1905.00426 [astro-ph.CO]}}.

\bibitem{2016A&A...596A.109P}
{Planck Collaboration}, N.~{Aghanim}, M.~{Ashdown}, {\em et~al.}, ``{Planck
  intermediate results. XLVIII. Disentangling Galactic dust emission and cosmic
  infrared background anisotropies},''
  \href{http://dx.doi.org/10.1051/0004-6361/201629022}{{\em \aap} {\bfseries
  596} (Dec., 2016) }.

\bibitem{Remazeilles11}
M.~{Remazeilles}, J.~{Delabrouille}, and J.-F. {Cardoso}, ``{Foreground
  component separation with generalized Internal Linear Combination},''
  \href{http://dx.doi.org/10.1111/j.1365-2966.2011.19497.x}{{\em \mnras}
  {\bfseries 418} (Nov., 2011) 467--476},
  \href{http://arxiv.org/abs/1103.1166}{{\ttfamily arXiv:1103.1166}}.

\bibitem{maniyar19}
A.~{Maniyar}, G.~{Lagache}, M.~{B{\'e}thermin}, and S.~{Ili{\'c}},
  ``{Constraining cosmology with the cosmic microwave and infrared backgrounds
  correlation},'' \href{http://dx.doi.org/10.1051/0004-6361/201833765}{{\em
  \aap} {\bfseries 621} (Jan, 2019) A32},
  \href{http://arxiv.org/abs/1809.04551}{{\ttfamily arXiv:1809.04551
  [astro-ph.CO]}}.

\bibitem{chon04}
G.~{Chon}, A.~{Challinor}, S.~{Prunet}, E.~{Hivon}, and I.~{Szapudi}, ``{Fast
  estimation of polarization power spectra using correlation functions},''
  \href{http://dx.doi.org/10.1111/j.1365-2966.2004.07737.x}{{\em \mnras}
  {\bfseries 350} no.~3, (May, 2004) 914--926},
  \href{http://arxiv.org/abs/astro-ph/0303414}{{\ttfamily
  arXiv:astro-ph/0303414 [astro-ph]}}.

\bibitem{planck2015cosmoparam}
{Planck Collaboration}, P.~A.~R. {Ade}, N.~{Aghanim}, {\em et~al.}, ``{Planck
  2015 results. XIII. Cosmological parameters},''
  \href{http://dx.doi.org/10.1051/0004-6361/201525830}{{\em \aap} {\bfseries
  594} (Sept., 2016) A13}, \href{http://arxiv.org/abs/1502.01589}{{\ttfamily
  arXiv:1502.01589 [astro-ph.CO]}}.

\bibitem{omori17}
Y.~{Omori}, R.~{Chown}, G.~{Simard}, {\em et~al.}, ``{A 2500 deg$^{2}$ CMB
  Lensing Map from Combined South Pole Telescope and Planck Data},''
  \href{http://dx.doi.org/10.3847/1538-4357/aa8d1d}{{\em \apj} {\bfseries 849}
  no.~2, (Nov, 2017) 124}, \href{http://arxiv.org/abs/1705.00743}{{\ttfamily
  arXiv:1705.00743 [astro-ph.CO]}}.

\bibitem{planck2018III}
{Planck Collaboration}, N.~{Aghanim}, Y.~{Akrami}, {\em et~al.}, ``{Planck 2018
  results. III. High Frequency Instrument data processing and frequency
  maps},'' {\em arXiv e-prints} (Jul, 2018) arXiv:1807.06207,
  \href{http://arxiv.org/abs/1807.06207}{{\ttfamily arXiv:1807.06207
  [astro-ph.CO]}}.

\bibitem{boehm16}
V.~{B{\"o}hm}, M.~{Schmittfull}, and B.~D. {Sherwin}, ``{Bias to CMB lensing
  measurements from the bispectrum of large-scale structure},''
  \href{http://dx.doi.org/10.1103/PhysRevD.94.043519}{{\em \prd} {\bfseries 94}
  no.~4, (Aug, 2016) 043519}, \href{http://arxiv.org/abs/1605.01392}{{\ttfamily
  arXiv:1605.01392 [astro-ph.CO]}}.

\bibitem{lewis2016}
A.~{Lewis} and G.~{Pratten}, ``{Effect of lensing non-Gaussianity on the CMB
  power spectra},'' \href{http://dx.doi.org/10.1088/1475-7516/2016/12/003}{{\em
  \jcap} {\bfseries 2016} no.~12, (Dec., 2016) 003},
  \href{http://arxiv.org/abs/1608.01263}{{\ttfamily arXiv:1608.01263
  [astro-ph.CO]}}.

\bibitem{planck2013XXX}
{Planck Collaboration}, P.~A.~R. {Ade}, N.~{Aghanim}, {\em et~al.}, ``{Planck
  2013 results. XXX. Cosmic infrared background measurements and implications
  for star formation},''
  \href{http://dx.doi.org/10.1051/0004-6361/201322093}{{\em \aap} {\bfseries
  571} (Nov, 2014) A30}, \href{http://arxiv.org/abs/1309.0382}{{\ttfamily
  arXiv:1309.0382 [astro-ph.CO]}}.

\bibitem{stein18}
G.~{Stein}, M.~A. {Alvarez}, and J.~R. {Bond}, ``{The mass-Peak Patch algorithm
  for fast generation of deep all-sky dark matter halo catalogues and its
  N-body validation},'' \href{http://dx.doi.org/10.1093/mnras/sty3226}{{\em
  \mnras} {\bfseries 483} no.~2, (Feb, 2019) 2236--2250},
  \href{http://arxiv.org/abs/1810.07727}{{\ttfamily arXiv:1810.07727
  [astro-ph.CO]}}.

\bibitem{stein20}
G.~{Stein}, M.~A. {Alvarez}, J.~R. {Bond}, A.~{van Engelen}, and
  N.~{Battaglia}, ``{The Websky Extragalactic CMB Simulations},'' {\em arXiv
  e-prints} (Jan., 2020) arXiv:2001.08787,
  \href{http://arxiv.org/abs/2001.08787}{{\ttfamily arXiv:2001.08787
  [astro-ph.CO]}}.

\bibitem{namikawa19}
T.~{Namikawa} and R.~{Takahashi}, ``{Impact of nonlinear growth of the
  large-scale structure on CMB B -mode delensing},''
  \href{http://dx.doi.org/10.1103/PhysRevD.99.023530}{{\em \prd} {\bfseries 99}
  no.~2, (Jan, 2019) 023530}, \href{http://arxiv.org/abs/1810.03346}{{\ttfamily
  arXiv:1810.03346 [astro-ph.CO]}}.

\bibitem{bk5}
{BICEP2 and Keck Array Collaborations}, P.~A.~R. {Ade}, Z.~{Ahmed}, {\em
  et~al.}, ``{BICEP2/Keck Array V: Measurements of B-mode Polarization at
  Degree Angular Scales and 150 GHz by the Keck Array},''
  \href{http://dx.doi.org/10.1088/0004-637X/811/2/126}{{\em \apj} {\bfseries
  811} no.~2, (Oct., 2015) 126},
  \href{http://arxiv.org/abs/1502.00643}{{\ttfamily arXiv:1502.00643
  [astro-ph.CO]}}.

\bibitem{planck2016SRoll}
{Planck Collaboration}, N.~{Aghanim}, M.~{Ashdown}, {\em et~al.}, ``{Planck
  intermediate results. XLVI. Reduction of large-scale systematic effects in
  HFI polarization maps and estimation of the reionization optical depth},''
  \href{http://dx.doi.org/10.1051/0004-6361/201628890}{{\em \aap} {\bfseries
  596} (Dec, 2016) A107}, \href{http://arxiv.org/abs/1605.02985}{{\ttfamily
  arXiv:1605.02985 [astro-ph.CO]}}.

\bibitem{cosmomc}
A.~Lewis and S.~Bridle, ``{Cosmological parameters from CMB and other data: a
  Monte- Carlo approach},'' {\em Phys. Rev.} {\bfseries D66} (2002) 103511,
\href{http://arxiv.org/abs/astro-ph/0205436}{{\ttfamily astro-ph/0205436}}.

\bibitem{biceparray20}
A.~{Schillaci}, P.~A.~R. {Ade}, Z.~{Ahmed}, {\em et~al.}, ``{Design and
  performance of the first BICEP Array receiver},'' {\em arXiv e-prints} (Feb.,
  2020) arXiv:2002.05228, \href{http://arxiv.org/abs/2002.05228}{{\ttfamily
  arXiv:2002.05228 [astro-ph.IM]}}.

\bibitem{bender18}
A.~N. {Bender}, P.~A.~R. {Ade}, Z.~{Ahmed}, {\em et~al.},
  \href{http://dx.doi.org/10.1117/12.2312426}{``{Year two instrument status of
  the SPT-3G cosmic microwave background receiver},''} in {\em Millimeter,
  Submillimeter, and Far-Infrared Detectors and Instrumentation for Astronomy
  IX}, vol.~10708 of {\em Society of Photo-Optical Instrumentation Engineers
  (SPIE) Conference Series}, p.~1070803.
\newblock Jul, 2018.
\newblock \href{http://arxiv.org/abs/1809.00036}{{\ttfamily arXiv:1809.00036
  [astro-ph.IM]}}.

\bibitem{SO18}
{The Simons Observatory Collaboration}, P.~{Ade}, J.~{Aguirre}, {\em et~al.},
  ``{The Simons Observatory: Science goals and forecasts},'' {\em ArXiv
  e-prints} (Aug., 2018) , \href{http://arxiv.org/abs/1808.07445}{{\ttfamily
  arXiv:1808.07445}}.

\bibitem{cmbs4-sb1}
{CMB-S4 Collaboration}, K.~N. {Abazajian}, P.~{Adshead}, {\em et~al.},
  ``{CMB-S4 Science Book, First Edition},'' {\em ArXiv e-prints} (Oct., 2016) ,
  \href{http://arxiv.org/abs/1610.02743}{{\ttfamily arXiv:1610.02743}}.

\bibitem{s4pgw}
{The CMB-S4 Collaboration}, K.~{Abazajian}, G.~E. {Addison}, {\em et~al.},
  ``{CMB-S4: Forecasting Constraints on Primordial Gravitational Waves},'' {\em
  arXiv e-prints} (Aug., 2020) arXiv:2008.12619,
  \href{http://arxiv.org/abs/2008.12619}{{\ttfamily arXiv:2008.12619
  [astro-ph.CO]}}.

\bibitem{pblens2019}
M.~A. {Fa{\'u}ndez}, K.~{Arnold}, C.~{Baccigalupi}, {\em et~al.},
  ``{Measurement of the Cosmic Microwave Background Polarization Lensing Power
  Spectrum from Two Years of POLARBEAR Data},''
  \href{http://dx.doi.org/10.3847/1538-4357/ab7e29}{{\em \apj} {\bfseries 893}
  no.~1, (Apr., 2020) 85}, \href{http://arxiv.org/abs/1911.10980}{{\ttfamily
  arXiv:1911.10980 [astro-ph.CO]}}.

\bibitem{actlens}
B.~D. {Sherwin}, A.~{van Engelen}, N.~{Sehgal}, {\em et~al.}, ``{Two-season
  Atacama Cosmology Telescope polarimeter lensing power spectrum},''
  \href{http://dx.doi.org/10.1103/PhysRevD.95.123529}{{\em \prd} {\bfseries 95}
  no.~12, (June, 2017) 123529},
  \href{http://arxiv.org/abs/1611.09753}{{\ttfamily arXiv:1611.09753
  [astro-ph.CO]}}.

\bibitem{manzotti18}
A.~{Manzotti}, ``{Future cosmic microwave background delensing with galaxy
  surveys},'' \href{http://dx.doi.org/10.1103/PhysRevD.97.043527}{{\em \prd}
  {\bfseries 97} no.~4, (Feb., 2018) 043527},
  \href{http://arxiv.org/abs/1710.11038}{{\ttfamily arXiv:1710.11038
  [astro-ph.CO]}}.

\bibitem{yu17}
B.~{Yu}, J.~C. {Hill}, and B.~D. {Sherwin}, ``{Multitracer CMB delensing maps
  from Planck and WISE data},''
  \href{http://dx.doi.org/10.1103/PhysRevD.96.123511}{{\em \prd} {\bfseries 96}
  no.~12, (Dec., 2017) 123511},
  \href{http://arxiv.org/abs/1705.02332}{{\ttfamily arXiv:1705.02332
  [astro-ph.CO]}}.

\bibitem{seljakhirata04}
U.~{Seljak} and C.~M. {Hirata}, ``{Gravitational lensing as a contaminant of
  the gravity wave signal in the CMB},''
  \href{http://dx.doi.org/10.1103/PhysRevD.69.043005}{{\em \prd} {\bfseries 69}
  no.~4, (Feb., 2004) 043005},
  \href{http://arxiv.org/abs/astro-ph/0310163}{{\ttfamily
  arXiv:astro-ph/0310163 [astro-ph]}}.

\bibitem{lensit}
J.~{Carron} and A.~{Lewis}, ``{Maximum a posteriori CMB lensing
  reconstruction},'' \href{http://dx.doi.org/10.1103/PhysRevD.96.063510}{{\em
  \prd} {\bfseries 96} no.~6, (Sept., 2017) 063510},
  \href{http://arxiv.org/abs/1704.08230}{{\ttfamily arXiv:1704.08230
  [astro-ph.CO]}}.

\bibitem{bayeslens}
M.~{Millea}, E.~{Anderes}, and B.~D. {Wandelt}, ``{Bayesian delensing delight:
  sampling-based inference of the primordial CMB and gravitational lensing},''
  {\em arXiv e-prints} (Feb., 2020) arXiv:2002.00965,
  \href{http://arxiv.org/abs/2002.00965}{{\ttfamily arXiv:2002.00965
  [astro-ph.CO]}}.

\bibitem{caldeira}
J.~{Caldeira}, W.~L.~K. {Wu}, B.~{Nord}, {\em et~al.}, ``{DeepCMB: Lensing
  reconstruction of the cosmic microwave background with deep neural
  networks},'' \href{http://dx.doi.org/10.1016/j.ascom.2019.100307}{{\em
  Astronomy and Computing} {\bfseries 28} (July, 2019) 100307},
  \href{http://arxiv.org/abs/1810.01483}{{\ttfamily arXiv:1810.01483
  [astro-ph.CO]}}.

\bibitem{ipython}
F.~P\'erez and B.~E. Granger, ``{IP}ython: a system for interactive scientific
  computing,'' \href{http://dx.doi.org/10.1109/MCSE.2007.53}{{\em Computing in
  Science and Engineering} {\bfseries 9} no.~3, (May, 2007) 21--29}.
  \url{https://ipython.org}.

\bibitem{matplotlib}
J.~D. Hunter, ``Matplotlib: A 2d graphics environment,''
  \href{http://dx.doi.org/10.1109/MCSE.2007.55}{{\em Computing in Science \&
  Engineering} {\bfseries 9} no.~3, (2007) 90--95}.

\bibitem{scipy}
P.~Virtanen, R.~Gommers, T.~E. Oliphant, {\em et~al.}, ``{{SciPy} 1.0:
  Fundamental Algorithms for Scientific Computing in Python},''
  \href{http://dx.doi.org/10.1038/s41592-019-0686-2}{{\em Nature Methods}
  {\bfseries 17} (2020) 261--272}.

\bibitem{healpy}
A.~Zonca, L.~Singer, D.~Lenz, {\em et~al.}, ``healpy: equal area pixelization
  and spherical harmonics transforms for data on the sphere in python,''
  \href{http://dx.doi.org/10.21105/joss.01298}{{\em Journal of Open Source
  Software} {\bfseries 4} no.~35, (Mar., 2019) 1298}.
  \url{https://doi.org/10.21105/joss.01298}.

\bibitem{healpix}
K.~M. {G{\'o}rski}, E.~{Hivon}, A.~J. {Banday}, {\em et~al.}, ``{HEALPix: A
  Framework for High-Resolution Discretization and Fast Analysis of Data
  Distributed on the Sphere},'' \href{http://dx.doi.org/10.1086/427976}{{\em
  \apj} {\bfseries 622} (Apr., 2005) 759--771},
  \href{http://arxiv.org/abs/arXiv:astro-ph/0409513}{{\ttfamily
  arXiv:astro-ph/0409513}}.

\end{thebibliography}\endgroup
\end{document}